\documentclass[twocolumn,preprintnumbers]{revtex4}
\usepackage{amssymb}
\usepackage{amsfonts}
\usepackage{amsmath}
\usepackage{graphicx}
\usepackage{bm}
\usepackage{hyperref}
\usepackage{version}

\input{tcilatex}
\begin{document}

\title{Wigner representation enables the exact derivation of the atom
interferometer phase, unlike the path integral approach.}
\author{B. Dubetsky}
\affiliation{Independent Researcher, 301 174th St, Sunny Isles Beach, Florida 33160 , USA}
\email{bdubetsky@gmail.com}
\date{\today }

\begin{abstract}
An exact expression for the phase of an atomic interferometer located in a
non-inertial reference frame (platform) moving along an arbitrary trajectory
and with an orientation that changes arbitrarily over time is obtained. This
expression takes into account precisely gravitational, Coriolis,
centrifugal, and gravity-gradient forces, which arise during the rotation of
the gravity source at a permanent rate. To achieve this result, we utilized
the equations for the atomic density matrix in the Wigner representation.
Starting from the exact formula, we derived three new terms in the
well-known limit of small rotation angles and short interrogation time,
which are attributed to the rotation and translational motion of the
platform.
\end{abstract}

\maketitle

%TCIMACRO{\TeXButton{onecolumngrid}{\onecolumngrid}}%
%BeginExpansion
\onecolumngrid%
%EndExpansion

\section{\label{s1}Introduction}

Since its birth about 40 years ago \cite{c1}, the field of atom
interferometry has matured significantly. The current state and prospects in
this area are presented, for example, in the reviews in \cite%
{c1.0,c1.1,c1.1.6,c1.1.7,c1.1.5,c1.1.4,c1.1.3,c1.1.2,c1.1.1,c1.1.0,c1.1.8},
and the proposals in \cite%
{c1.2,c1.3,c1.4,c1.5,c1.5.1,c1.5.2,c1.5.3,c1.5.6,c1.5.7,c1.5.5,c1.5.9,c1.5.4,c1.5.8,c1.5.10}%
.

Atom interferometry is a phenomenon caused by the quantization of the atomic
center-of-mass motion. Since \cite{c2}, density matrix equations in the
Wigner representation traditionally form the theoretical basis for these
phenomena \cite{c3,c4}. For example, these equations are widely used in the
theory of laser cooling of atoms \cite{c5}. However, one can mention only
two examples when this traditional approach was used for atomic
interferometers: these are \cite{c6} and the subsequent papers \cite{c7,c8},
and \cite{c9}. Since \cite{c10}, authors use the path integral approach \cite%
{c11} much more frequently. Examples of such use one can find in articles 
\cite{c12,c12.1,c13,c14,c15,c16,c16.1} and in dissertations \cite%
{c17,c18,c19,c20,c21,c22}. Despite wide distribution, path integral approach
results are often approximate, not exact. As for the phase of the atomic
interferometer (AI), only acceleration, gravitational and inertial, is taken
into account exactly \cite{c12}. The remaining contributions, due to such
effects as the Earth's rotation \cite{c12}, first-order gravity-gradients
and quantum corrections \cite{c12.1}, higher-order gravity-gradients \cite%
{c23,c7,c23.1,c24,c25,c26}, were considered as small perturbations. Of
course, all these results were also obtained using the equations in the
Wigner representation \cite{c6,c7}. For the above-listed approximate
results, one used the expansion of the classical trajectory of the atom in
powers of the interrogation time $T$. In the work \cite{c27} the first 7
terms of such an expansion were obtained. In the important case when the
atom moves in the field of a gravitational source rotating with a permanent
rate and one takes into account only the first-order gravity-gradient, the
trajectory of the atom obeys a linear equation with time-independent
coefficients and riht-hand-side. As a result, it is possible to derive an
exact expression for the trajectory of the atom. Using the approach
developed in \cite{c28}, such an exact classical trajectory of the atom was
obtained in \cite{c29,c30,c31}. In \cite{c29,c30}, these trajectories served
to calculate the AI phase, but a calculation made without acknowledging the
quantization of the atom's center-of-mass motion. Article \cite{c27}
provides another example of deriving and using an atom's exact classical
trajectory. However, the gravitational source's rotation was not included
there.

In the present work, using the equation in the Wigner representation, I
obtained an exact analytical expression for the phase of the AI in a
non-inertial frame (platform) moving along an arbitrary trajectory with an
orientation that arbitrarily changes in time relative to the gravitational
source rotating with a permanent rate. In this phase, in addition to the
exact accounting of inertial effects associated with the motion and rotation
of the platform, I took into account the 1st order gravity-gradient forces,
Coriolis and centrifugal forces due to the rotation of the gravity source. I
was unable to obtain this result using the path integral technique. Should
this phase expression elude one in future path integral calculations - and
it might - one should have to concede that Wigner representation equations
reign supreme in the atomic interferometry.

The paper is organized as follows. In the next section, one considers the
evolution of the atomic density matrix in the Wigner representation in the
time between the field pulses. The phase of the AI is calculated in Sec. \ref%
{s3}. The case of small values of the rotation rate and the gravity-gradient
tensor of the gravity source, as well as small rotation of the platform, is
considered in Sec. \ref{s4}. The results and possible applications are
discussed in Sec. \ref{s5}.

\section{\label{s2}Density matrix}

Calculating phase in the platform system requires a Lagrangian (path
integrals) or Hamiltonian (Wigner representation). The peculiarity of our
problem is that the platform moves relative to the non-inertial frame of the
rotating gravity source (RGS). The advantage of this frame is that it is the
only one where gravity is time-independent. With no textbook references
available, this case requires consideration here.

The Lagrangian in the RGS-frame is\ given by \cite{c32}

\begin{equation}
L=M\left[ \dfrac{\mathbf{V}_{E}^{2}}{2}+\mathbf{V}_{E}\cdot \left( \mathbf{%
\Omega }_{E}\times \mathbf{X}_{E}\right) +\dfrac{\left( \mathbf{\Omega }%
_{E}\times \mathbf{X}_{E}\right) ^{2}}{2}-U\left( \mathbf{X}_{E}\right) %
\right] ,  \label{1}
\end{equation}%
where $M,\mathbf{V}_{E}$ and $\mathbf{X}_{E}$ are the mass, velocity and
coordinate of the atom, $U\left( \mathbf{X}_{E}\right) $ is the potential of
the gravitational field.

In the platform system,

\begin{subequations}
\label{2}
\begin{eqnarray}
\mathbf{x} &=&\underline{R}\left( \mathbf{X}_{E}-\mathbf{X}\right) ,
\label{2a} \\
\mathbf{v} &=&-\mathbf{\Omega }\times \mathbf{x}+\underline{R}\left( \mathbf{%
V}_{E}-\mathbf{V}\right) ,  \label{2b}
\end{eqnarray}%
where $\mathbf{X},\mathbf{V},$ \underline{$R$} and \textbf{$\Omega $} are
the coordinate, velocity, rotation matrix and rotation rate of the platform,
one gets

\end{subequations}
\begin{equation}
L=M\left[ \dfrac{\left[ \underline{R}^{T}\left( \mathbf{v}+\mathbf{\Omega }%
\times \mathbf{x}\right) +\mathbf{V}\right] ^{2}}{2}+\left[ \underline{R}%
^{T}\left( \mathbf{v}+\mathbf{\Omega }\times \mathbf{x}\right) +\mathbf{V}%
\right] \cdot \left[ \mathbf{\Omega }_{E}\times \left( \underline{R}^{T}%
\mathbf{x}+\mathbf{X}\right) \right] +\dfrac{\left( \mathbf{\Omega }%
_{E}\times \left( \underline{R}^{T}\mathbf{x}+\mathbf{X}\right) \right) ^{2}%
}{2}-U\left( \mathbf{X}_{E}\right) \right] .  \label{3}
\end{equation}%
Let us now assume that the atom moves in a small neighborhood of the point $%
\mathbf{X}_{c}$,%
\begin{equation}
\mathbf{X}_{E}=\mathbf{x}_{E}+\mathbf{X}_{c},  \label{4}
\end{equation}%
where one keeps only zero, first, and second order terms in the potential
expansion over $\mathbf{x}_{E},$

\begin{equation}
U\left( \mathbf{X}_{E}\right) \approx U\left( \mathbf{X}_{c}\right) -M\left[ 
\mathbf{g}_{E}\left( \mathbf{X}_{c}\right) \cdot \mathbf{x}_{E}+\dfrac{1}{2}%
\mathbf{x}_{E}^{T}\underline{\mathbf{\gamma }}_{E}\left( \mathbf{X}%
_{c}\right) \mathbf{x}_{E}\right] ,  \label{5}
\end{equation}%
$\mathbf{g}_{E}\left( \mathbf{X}_{c}\right) $ and \underline{$\mathbf{\gamma 
}$}$_{E}\left( \mathbf{X}_{c}\right) $ are the gravitational acceleration
and the first-order gravity-gradient tensor. Then one gets for Lagrangian (%
\ref{3})

\begin{subequations}
\label{6}
\begin{align}
L& =M\left\{ \dfrac{v^{2}}{2}+\mathbf{v\cdot }\left[ \left( \mathbf{\Omega }%
\left( t\right) +\underline{R}\left( t\right) \mathbf{\Omega }_{E}\right)
\times \mathbf{x}\right] +\dfrac{\left( \mathbf{\Omega }\left( t\right)
\times \mathbf{x}\right) \cdot \left[ \left( \mathbf{\Omega }\left( t\right)
+2\underline{R}\left( t\right) \mathbf{\Omega }_{E}\right) \times \mathbf{x}%
\right] }{2}+\mathbf{x}\cdot \mathbf{a}+\dfrac{1}{2}\mathbf{x}\underline{R}%
\left( t\right) \underline{\mathbf{\gamma }}\left( \mathbf{X}_{c}\right) 
\underline{R}^{T}\left( t\right) \mathbf{x}\right\} ,  \label{6a} \\
& \mathbf{a}=\underline{R}\left( t\right) \left[ \mathbf{g}\left( \mathbf{X}%
_{c}\right) -\mathbf{A}\left( t\right) -2\mathbf{\Omega }_{E}\times \mathbf{V%
}+\underline{\mathbf{\gamma }}\left( \mathbf{X}_{c}\right) \left( \mathbf{X}%
\left( t\right) -\mathbf{X}_{c}\right) \right] ,  \label{6b} \\
\mathbf{g}\left( \mathbf{X}_{c}\right) & =\mathbf{g}_{E}\left( \mathbf{X}%
_{c}\right) -\mathbf{\Omega }_{E}\times \left( \mathbf{\Omega }_{E}\times 
\mathbf{X}_{c}\right) ,  \label{6c} \\
\underline{\mathbf{\gamma }}\left( \mathbf{X}_{c}\right) & =\underline{%
\mathbf{\gamma }}_{E}\left( \mathbf{X}_{c}\right) +\mathbf{\Omega }_{E}^{2}%
\underline{I}-\mathbf{\Omega }_{E}\mathbf{\Omega }_{E}^{T},  \label{6d}
\end{align}%
where \textbf{$A$}$\left( t\right) $ is the acceleration of the platform and 
\underline{$I$} is the unity matrix. From here one comes to the navigation
equations

\end{subequations}
\begin{equation}
\overset{_{\bullet \bullet }}{\mathbf{x}}=\vec{a}-\overset{_{\bullet }}{%
\mathbf{\Omega }}\times \mathbf{x}-2\left( \mathbf{\Omega }+\underline{R}%
\mathbf{\Omega }_{E}\right) \times \mathbf{v}-\left( \mathbf{\Omega }+2%
\underline{R}\mathbf{\Omega }_{E}\right) \times \left( \mathbf{\Omega }%
\times \mathbf{x}\right) +\underline{R}\underline{\mathbf{\gamma }}\left( 
\mathbf{X}_{c}\right) \underline{R}^{T}\mathbf{x}  \label{6.1}
\end{equation}%
To use the path integral method, it is necessary to substitute expressions
for the position and velocity of the atom $\left\{ \mathbf{x},\mathbf{v}%
\right\} $ into this Lagrangian, and they, as can be seen from Eqs. (\ref{2}%
, \ref{4}) depend not only on the given trajectory and orientation of the
platform, but also on the trajectory of the atom in the RGS frame $\mathbf{x}%
_{E}$, for which the following expressions are derived \cite{c31} 
\begin{subequations}
\label{7}
\begin{eqnarray}
\mathbf{x}_{E}\left( t\right) &=&\mathbf{\xi }_{E}\left[ \mathbf{x}%
_{E}\left( t^{\prime }\right) ,\mathbf{V}_{E}\left( t^{\prime }\right)
,t-t^{\prime }\right] ,  \label{7a} \\
\mathbf{\xi }_{E}\left( \mathbf{x},\mathbf{v},t\right) &=&\underline{x}%
_{x}\left( t\right) \mathbf{x}+\underline{x}_{v}\left( t\right) \mathbf{v}+%
\underline{x}_{g}\left( t\right) \mathbf{g}\left( \mathbf{X}_{c}\right) ,
\label{7b} \\
\underline{x}_{x}\left( t\right) &=&\left[ \underline{x}_{+}\exp \left( -i%
\underline{\tilde{\omega}}t\right) +\underline{x}_{-}\exp \left( i\underline{%
\tilde{\omega}}t\right) \underline{\tilde{\omega}}^{-1}\underline{x}_{-}^{-1}%
\underline{x}_{+}\underline{\tilde{\omega}}\right] \left( \underline{x}_{+}+%
\underline{x}_{-}\underline{\tilde{\omega}}^{-1}\underline{x}_{-}^{-1}%
\underline{x}_{+}\underline{\tilde{\omega}}\right) ^{-1},  \label{7c} \\
\underline{x}_{v}\left( t\right) &=&i\left( \underline{x}_{+}\exp \left( -i%
\underline{\tilde{\omega}}t\right) \underline{x}_{+}^{-1}\underline{x}_{-}-%
\underline{x}_{-}\exp \left( i\underline{\tilde{\omega}}t\right) \right)
\left( \underline{x}_{-}\underline{\tilde{\omega}}+\underline{x}_{+}%
\underline{\tilde{\omega}}\underline{x}_{+}^{-1}\underline{x}_{-}\right)
^{-1},  \label{7d} \\
\underline{x}_{g}\left( t\right) &=&\left[ \underline{x}_{+}\left( I-\exp
\left( -i\underline{\tilde{\omega}}t\right) \right) \underline{\tilde{\omega}%
}^{-1}\underline{x}_{+}^{-1}\underline{x}_{-}+\underline{x}_{-}\left( 1-\exp
\left( i\underline{\tilde{\omega}}t\right) \right) \underline{\tilde{\omega}}%
^{-1}\right] \left( \underline{x}_{-}\underline{\tilde{\omega}}+\underline{x}%
_{+}\underline{\tilde{\omega}}\underline{x}_{+}^{-1}\underline{x}_{-}\right)
^{-1},  \label{7e} \\
\underline{x}_{\pm } &\equiv &\left\{ \mathbf{x}\left( \pm \omega
_{1}\right) ,\mathbf{x}\left( \pm \omega _{2}\right) ,\mathbf{x}\left( \pm
\omega _{3}\right) \right\} ,  \label{7f} \\
\tilde{\omega} &=&diag\left\{ \omega _{1},\omega _{2},\omega _{3}\right\}
\label{7g} \\
\mathbf{x}\left( \omega \right) &=&\left( 
\begin{array}{c}
\underline{A}_{22}\left( \omega \right) \underline{A}_{33}\left( \omega
\right) -\underline{A}_{23}\left( \omega \right) \underline{A}_{32}\left(
\omega \right) \\ 
\underline{A}_{23}\left( \omega \right) \underline{A}_{31}\left( \omega
\right) -\underline{A}_{21}\left( \omega \right) \underline{A}_{33}\left(
\omega \right) \\ 
\underline{A}_{21}\left( \omega \right) \underline{A}_{32}\left( \omega
\right) -\underline{A}_{22}\left( \omega \right) \underline{A}_{31}\left(
\omega \right)%
\end{array}%
\right) ,  \label{7h} \\
\underline{A}\left( \omega \right) &=&\underline{\mathbf{\gamma }}\left(
X_{c}\right) +\omega ^{2}\underline{I}+2\omega \underline{\Omega }_{E},
\label{7i} \\
\underline{\Omega }_{ik} &=&-\varepsilon _{ikl}\mathbf{\Omega }_{l},
\label{7j}
\end{eqnarray}%
where for the eigenfrequencies one has 
\end{subequations}
\begin{equation}
\omega _{m}=\sqrt{z_{m}-\dfrac{a_{2}}{3}},  \label{8}
\end{equation}%
where $z_{m}$ are the well-known roots of the equation

\begin{subequations}
\label{9}
\begin{eqnarray}
z^{3}+pz+s &=&0,  \label{9a} \\
p &=&\ a_{1}-\dfrac{a_{2}^{2}}{3},  \label{9b} \\
s &=&a_{0}-\dfrac{a_{2}a_{1}}{3}+\dfrac{2a_{2}^{3}}{27},  \label{9c} \\
a_{0} &=&\left\vert \underline{\mathbf{\gamma }}\left( X_{c}\right)
\right\vert ,  \label{9d} \\
a_{1} &=&\dfrac{1}{2}\left( Tr^{2}\left( \underline{\mathbf{\gamma }}\left(
X_{c}\right) \right) -Tr\left( \underline{\mathbf{\gamma }}^{2}\left(
X_{c}\right) \right) \right) -4\mathbf{\Omega }_{E}\cdot \underline{\mathbf{%
\gamma }}\left( X_{c}\right) \mathbf{\Omega }_{E},  \label{9e} \\
a_{2} &=&Tr\left( \underline{\mathbf{\gamma }}\left( X_{c}\right) \right) -4%
\mathbf{\Omega }_{E}^{2}.  \label{9f}
\end{eqnarray}%
The matrices \underline{$x$}$_{x}=A$ and \underline{$x$}$_{v}=B$ together
make up the first row $\left\{ A,B\right\} $ in the ABCD matrix, according
to Ref. \cite{c31.1}. Row $\left\{ A,B\right\} $ was calculated for the
motion along vertical $z-$axis using \underline{$\gamma $}$_{zz}$-component
of the gravity gradient tensor $\gamma _{E}\left( \mathbf{X}_{c}\right) .$%
The generalization (\ref{7c}, \ref{7d}) for arbitrary tensor $\gamma
_{E}\left( \mathbf{X}_{c}\right) $ with gravity source rotating at arbitrary
permanent rate $\mathbf{\Omega }_{E}$ was obtained in Ref. \cite{c31}.

The expressions for the Lagrangian (\ref{6a}) and the atomic trajectory (\ref%
{2}, \ref{4}, \ref{7}-\ref{9}) are sufficient to calculate the AI phase
using the path integral technique. Let us now turn to the Wigner
representation. The atomic momentum is 
\end{subequations}
\begin{equation}
\mathbf{p}=M\left\{ \mathbf{v}+\left[ \left( \mathbf{\Omega }+\underline{R}%
\mathbf{\Omega }_{E}\right) \times \mathbf{x}\right] \right\} .  \label{10}
\end{equation}%
From here and from Eq. (\ref{6}) one gets for the Hamiltonian%
\begin{equation}
H=\dfrac{\mathbf{p}^{2}}{2M}-\mathbf{p\cdot }\left[ \left( \mathbf{\Omega }+%
\underline{R}\mathbf{\Omega }_{E}\right) \times \mathbf{x}\right] -M\left[ 
\mathbf{a\cdot x}+\dfrac{1}{2}\mathbf{x}^{T}\underline{R}\underline{\mathbf{%
\gamma }}_{E}\left( \mathbf{X}_{c}\right) \underline{R}^{T}\mathbf{x}\right]
.  \label{11}
\end{equation}%
Except for taking into account the rotation and motion of the platform, and
the fact that one here considers only the gravitational potential (\ref{5}),
Hamiltonian (\ref{11}) has the same structure as Hamiltonian (12b) in \cite%
{c6}. Therefore, when one redoes the math from that article, one can see
that the quantum density matrix in the Wigner representation%
\begin{equation}
\rho \left( \mathbf{x},\mathbf{p},t\right) =\int \dfrac{d\vec{s}}{\left(
2\pi \hbar \right) ^{3}}\rho \left( \mathbf{x}+\dfrac{1}{2}\mathbf{s},%
\mathbf{x}-\dfrac{1}{2}\mathbf{s},t\right) \exp \left( -i\mathbf{p}\cdot 
\mathbf{s}/\hbar \right) ,  \label{12}
\end{equation}%
where

\begin{equation}
\rho \left( \mathbf{x},\mathbf{x}^{\prime },t\right) =\psi \left( \mathbf{x}%
,t\right) \otimes \psi \dag \left( \mathbf{x}^{\prime },t\right)  \label{13}
\end{equation}%
is the density matrix in the coordinate representation, obeys the exact same
Liouville equation as the classical counterpart during the time between
field pulses.%
\begin{equation}
\left\{ \partial _{t}+\overset{_{\bullet }}{\mathbf{x}}\cdot \partial _{%
\mathbf{x}}+\overset{_{\bullet }}{\mathbf{p}}\cdot \partial _{\mathbf{p}%
}\right\} \rho \left( \mathbf{x},\mathbf{p},t\right) =0,  \label{14}
\end{equation}%
where

\begin{subequations}
\label{15}
\begin{eqnarray}
\overset{_{\bullet }}{\mathbf{x}} &=&\dfrac{\mathbf{p}}{M}-\left( \mathbf{%
\Omega }+\underline{R}\mathbf{\Omega }_{E}\right) \times \mathbf{x},
\label{15a} \\
\overset{_{\bullet }}{\mathbf{p}} &=&-\left( \mathbf{\Omega }+\underline{R}%
\mathbf{\Omega }_{E}\right) \times \mathbf{p}+M\left( \mathbf{a}+\underline{R%
}\underline{\mathbf{\gamma }}_{E}\left( \mathbf{X}_{c}\right) \underline{R}%
^{T}\mathbf{x}\right) .  \label{15b}
\end{eqnarray}%
From the conservation of the number of particles in the phase space $\left\{ 
\mathbf{x},\mathbf{p}\right\} $ it follows that 
\end{subequations}
\begin{equation}
\rho \left( \mathbf{x},\mathbf{p},t\right) =\rho \left[ \mathbf{\xi }\left( 
\mathbf{x},\mathbf{p},t^{\prime },t\right) ,\mathbf{\pi }\left( \mathbf{x},%
\mathbf{p},t^{\prime },t\right) ,t^{\prime }\right] ,  \label{16}
\end{equation}%
where $\mathbf{\xi }\left( \mathbf{x},\mathbf{p},t^{\prime },t\right) $ and $%
\mathbf{\pi }\left( \mathbf{x},\mathbf{p},t^{\prime },t\right) $ are the
Eqs. (\ref{15}) propagators \cite{c32.1}, i.e. the solutions of those
equations are 
\begin{subequations}
\label{17}
\begin{eqnarray}
\mathbf{x}\left( t\right) &=&\mathbf{\xi }\left[ \mathbf{x}\left( t^{\prime
}\right) ,\mathbf{p}\left( t^{\prime }\right) ,t,t^{\prime }\right] ,
\label{17a} \\
\mathbf{p}\left( t\right) &=&\mathbf{\pi }\left[ \mathbf{x}\left( t^{\prime
}\right) ,\mathbf{p}\left( t^{\prime }\right) ,t,t^{\prime }\right] .
\label{17b}
\end{eqnarray}%
Using Eqs. (\ref{2}, \ref{4}, \ref{7}, \ref{10}), one gets for propagators 
\end{subequations}
\begin{subequations}
\label{18}
\begin{eqnarray}
&&\mathbf{\xi }\left( \mathbf{x},\mathbf{p},t,t^{\prime }\right) =\underline{%
R}\left( t\right) \left\{ \left[ \underline{x}_{x}\left( t-t^{\prime
}\right) -\underline{x}_{v}\left( t-t^{\prime }\right) \underline{\Omega }%
_{E}\right] \underline{R}^{T}\left( t^{\prime }\right) \mathbf{x}+\underline{%
x}_{v}\left( t-t^{\prime }\right) \left[ \underline{R}^{T}\left( t^{\prime
}\right) \dfrac{\mathbf{p}}{M}+\mathbf{V}\left( t^{\prime }\right) \right]
\right.  \notag \\
&&\left. +\underline{x}_{g}\left( t-t^{\prime }\right) \mathbf{g}\left( 
\mathbf{X}_{c}\right) +\underline{x}_{x}\left( t-t^{\prime }\right) \left[ 
\mathbf{X}\left( t^{\prime }\right) -\mathbf{X}_{c}\right] -\mathbf{X}\left(
t\right) +\mathbf{X}_{c}\right\} ,  \label{18a} \\
&&\mathbf{\pi }\left( \mathbf{x},\mathbf{p},t,t^{\prime }\right) =M%
\underline{R}\left( t\right) \left\{ \left[ \underline{\dot{x}}_{x}\left(
t-t^{\prime }\right) -\underline{\dot{x}}_{v}\left( t-t^{\prime }\right) 
\underline{\Omega }_{E}+\underline{\Omega }_{E}\left( \underline{x}%
_{x}\left( t-t^{\prime }\right) -\underline{x}_{v}\left( t-t^{\prime
}\right) \underline{\Omega _{E}}\right) \right] \underline{R}^{T}\left(
t^{\prime }\right) \mathbf{x}\right.  \notag \\
&&+\left[ \underline{\dot{x}}_{v}\left( t-t^{\prime }\right) +\underline{%
\Omega }_{E}\underline{x}_{v}\left( t-t^{\prime }\right) \right] \left[ 
\underline{R}^{T}\left( t^{\prime }\right) \dfrac{\mathbf{p}}{M}+\mathbf{V}%
\left( t^{\prime }\right) \right] +\left[ \underline{\dot{x}}_{g}\left(
t-t^{\prime }\right) +\underline{\Omega }_{E}\underline{x}_{g}\left(
t-t^{\prime }\right) \right] \mathbf{g}\left( \mathbf{X}_{c}\right)  \notag
\\
&&\left. +\left[ \underline{\dot{x}}_{x}\left( t-t^{\prime }\right) +%
\underline{\Omega }_{E}\underline{x}_{x}\left( t-t^{\prime }\right) \right] %
\left[ \mathbf{X}\left( t^{\prime }\right) -\mathbf{X}_{c}\right] -%
\underline{\Omega }_{E}\left[ \mathbf{X}\left( t\right) -\mathbf{X}_{c}%
\right] -\mathbf{V}\left( t\right) \right\} .  \label{18b}
\end{eqnarray}%
The propagators satisfy the multiplication law 
\end{subequations}
\begin{equation}
\left( 
\begin{array}{c}
\mathbf{\xi } \\ 
\mathbf{\pi }%
\end{array}%
\right) \left[ \mathbf{\xi }\left( \mathbf{x},\mathbf{p},t_{2},t_{1}\right) ,%
\mathbf{\pi }\left( \mathbf{x},\mathbf{p},t_{2},t_{1}\right) ,t_{3},t_{2}%
\right] =\left( 
\begin{array}{c}
\mathbf{\xi } \\ 
\mathbf{\pi }%
\end{array}%
\right) \left( \mathbf{x},\mathbf{p},t_{3},t_{1}\right) .  \label{18.1}
\end{equation}%
To verify the solution (\ref{16}) one must substitute this solution into the
Liouville equation (\ref{14}) and obtain an identity. In the paper \cite{c6}
for a stationary and non-rotating platform the authors proved the identity
using the invariance of the Poisson brackets and the fact that the
propagators depend only on the difference in times $t$ and $t^{\prime }.$
Obviously, when the platform moves and rotates, the propagators (\ref{18})
depend separately on the final and initial times $t$ and $t^{\prime }$ and
the proof found in \cite{c6} does not work. However, it is obvious that it
is enough to consider the case when the initial density matrix is

\begin{equation}
\rho \left( \mathbf{x},\mathbf{p},t^{\prime }\right) =\delta \left( \mathbf{x%
}-\mathbf{x}_{0}\right) \delta \left( \mathbf{p}-\mathbf{p}_{0}\right) .
\label{19}
\end{equation}%
Then one will find that the matrices \underline{$x$}$_{x},$ \underline{$x$}$%
_{v},$ \underline{$x$}$_{g}$ from Eqs. (\ref{7}) and their derivatives must
satisfy the following equations 
\begin{subequations}
\label{20}
\begin{gather}
-\underline{\dot{x}}_{x}\left( t^{\prime }-t\right) +\left[ \underline{\dot{x%
}}_{v}\left( t^{\prime }-t\right) -\underline{x}_{x}\left( t^{\prime
}-t\right) \right] \underline{\Omega }_{E}+\underline{x}_{v}\left( t^{\prime
}-t\right) \left[ \underline{\mathbf{\gamma }}\left( \mathbf{X}_{c}\right) +2%
\underline{\Omega }_{E}^{2}\right] =0,  \label{20a} \\
-\underline{\dot{x}}_{v}\left( t^{\prime }-t\right) +\underline{x}_{x}\left(
t^{\prime }-t\right) -2\underline{x}_{v}\left( t^{\prime }-t\right) 
\underline{\Omega }_{E}=0,  \label{20b} \\
-\underline{\dot{x}}_{x}\left( t^{\prime }-t\right) +\underline{x}_{v}\left(
t^{\prime }-t\right) \underline{\mathbf{\gamma }}\left( \mathbf{X}%
_{c}\right) =0,  \label{20c} \\
-\underline{\ddot{x}}_{x}\left( t^{\prime }-t\right) +\underline{\ddot{x}}%
_{v}\left( t^{\prime }-t\right) \underline{\Omega }_{E}+\underline{\Omega }%
_{E}\left( -\underline{\dot{x}}_{x}\left( t^{\prime }-t\right) +\underline{%
\dot{x}}_{v}\left( t^{\prime }-t\right) \underline{\Omega }_{E}\right) =0,
\label{20d} \\
-\left( \underline{\dot{x}}_{x}\left( t^{\prime }-t\right) +\underline{%
\Omega }_{E}\underline{x}_{x}\left( t^{\prime }-t\right) \right) \underline{%
\Omega }_{E}+\left( \underline{\dot{x}}_{v}\left( t^{\prime }-t\right) +%
\underline{\Omega }_{E}\underline{x}_{v}\left( t^{\prime }-t\right) \right)
\left( \underline{\mathbf{\gamma }}\left( \mathbf{X}_{c}\right) +2\underline{%
\Omega }_{E}^{2}\right) =0,  \label{20e} \\
-\underline{\ddot{x}}_{v}\left( t^{\prime }-t\right) -\underline{\Omega }_{E}%
\underline{\dot{x}}_{v}\left( t^{\prime }-t\right) +\underline{\dot{x}}%
_{x}\left( t^{\prime }-t\right) +\underline{\Omega }_{E}\underline{x}%
_{x}\left( t^{\prime }-t\right) -2\left( \underline{\dot{x}}_{v}\left(
t^{\prime }-t\right) +\underline{\Omega }_{E}\underline{x}_{v}\left(
t^{\prime }-t\right) \right) \underline{\Omega }_{E}=0,  \label{20f} \\
-\underline{\ddot{x}}_{g}\left( t^{\prime }-t\right) -\underline{\Omega }_{E}%
\underline{\dot{x}}_{g}\left( t^{\prime }-t\right) +\underline{\dot{x}}%
_{v}\left( t^{\prime }-t\right) +\underline{\Omega }_{E}\underline{x}%
_{v}\left( t^{\prime }-t\right) =0,  \label{20g} \\
-\underline{\ddot{x}}_{x}\left( t^{\prime }-t\right) -\underline{\Omega }_{E}%
\underline{\dot{x}}_{x}\left( t^{\prime }-t\right) +\left( \underline{\dot{x}%
}_{v}\left( t^{\prime }-t\right) +\underline{\Omega }_{E}\underline{x}%
_{v}\left( t^{\prime }-t\right) \right) \underline{\mathbf{\gamma }}\left( 
\mathbf{X}_{c}\right) =0,  \label{20h} \\
-\underline{\dot{x}}_{g}\left( t^{\prime }-t\right) +\underline{x}_{v}\left(
t^{\prime }-t\right) =0.  \label{20i}
\end{gather}

Only the last identity obviously follows from Eqs. (\ref{7d}, \ref{7e}). We
were unable to prove analytically that the remaining equations (\ref{20})
are identities. We, however, verified them numerically, randomly choosing
time $t^{\prime }-t<0,$ tensor \underline{$\mathbf{\gamma }$}$\left( \mathbf{%
X}_{c}\right) $ and vector \textbf{$\Omega $}$_{E}$. If we define the norm
of matrix \underline{$a$} as

\end{subequations}
\begin{equation}
\left\Vert \underline{a}\right\Vert \equiv \dfrac{1}{3}\sqrt{%
\dsum_{i=1}^{3}\dsum_{k=1}^{3}\underline{a}_{ik}^{2}},  \label{21}
\end{equation}%
then after $15000$ samples the norms of matrices on the left-hand side of
Eqs. (\ref{20}) do not exceed $1.7\times 10^{-11}$. These tests, could lead
us to believe that the analytical expressions (\ref{18}) for propagators
stand on solid ground.

\section{\label{s3}Exact phase}

Since \cite{c33}, Raman field pulses resonant to the transition between two
sublevels of the ground state of an atom, which one calls ground $\left\vert
g\right\rangle =\left( 
\begin{array}{c}
0 \\ 
1%
\end{array}%
\right) $ and excited $\left\vert e\right\rangle =\left( 
\begin{array}{c}
1 \\ 
0%
\end{array}%
\right) $ are often used for AI. If we neglect the displacement and
acceleration of the atom under the action of gravitational and inertial
forces during the pulse, then the action of the field pulse leads to a
quantum jump of the density matrix. These jumps were calculated in \cite{c6}
for an arbitrary pulse area. However, here one considers only the sequence
of $\pi /2-\pi -\pi /2$ pulses acting at the moments $T_{1}.$ $T_{2}$ and $%
T_{3}$. For $\pi /2-$pulse the quantum jump is 
\begin{subequations}
\label{22}
\begin{eqnarray}
\rho _{ee}^{\left( +\right) }\left( \mathbf{x},\mathbf{p},t\right) &=&\dfrac{%
1}{2}\left[ \rho _{ee}^{\left( -\right) }\left( \mathbf{x},\mathbf{p}%
,t\right) +\rho _{gg}^{\left( -\right) }\left( \mathbf{x},\mathbf{p}-\hbar 
\mathbf{k},t\right) \right] +\func{Re}\left\{ i\exp \left[ -i\left( \mathbf{k%
}\cdot \mathbf{x}-\delta t-\phi \right) \right] \rho _{eg}^{\left( -\right)
}\left( \mathbf{x},\mathbf{p}-\dfrac{\hbar \mathbf{k}}{2},t\right) \right\} ,
\label{22a} \\
\rho _{gg}^{\left( +\right) }\left( \mathbf{x},\mathbf{p},t\right) &=&\dfrac{%
1}{2}\left[ \rho _{ee}^{\left( -\right) }\left( \mathbf{x},\mathbf{p}+\hbar 
\mathbf{k},t\right) +\rho _{gg}^{\left( -\right) }\left( \mathbf{x},\mathbf{p%
},t\right) \right] +\func{Re}\left\{ i\exp \left[ -i\left( \mathbf{k}\cdot 
\mathbf{x}-\delta t-\phi \right) \right] \rho _{eg}^{\left( -\right) }\left( 
\mathbf{x},\mathbf{p}+\dfrac{\hbar \mathbf{k}}{2},t\right) \right\} ,
\label{22b} \\
\rho _{eg}^{\left( +\right) }\left( \mathbf{x},\mathbf{p},t_{+}\right) &=&%
\dfrac{i}{2}\exp \left[ i\left( \mathbf{k}\cdot \mathbf{x}-\delta t-\phi
\right) \right] \left[ \rho _{ee}^{\left( -\right) }\left( \mathbf{x},%
\mathbf{p}+\dfrac{\hbar \mathbf{k}}{2},t\right) -\rho _{gg}^{\left( -\right)
}\left( \mathbf{x},\mathbf{p}-\dfrac{\hbar \mathbf{k}}{2},t\right) \right] 
\notag \\
&&+\dfrac{1}{2}\left\{ \rho _{eg}^{\left( -\right) }\left( \mathbf{x},%
\mathbf{p},t\right) +\exp \left[ 2i\left( \mathbf{k}\cdot \mathbf{x}-\delta
t-\phi \right) \right] \rho _{ge}^{\left( -\right) }\left( \mathbf{x},%
\mathbf{p},t\right) \right\} ,  \label{22c}
\end{eqnarray}%
where $\rho ^{\left( \pm \right) }$ are the density matrices after and
before the pulse, $\mathbf{k},$ $\phi ,$ and $\delta $ are the effective
wave vector, phase, and frequency detuning of the Raman field from the $%
e\rightarrow g$ atomic transition frequency. For $\pi -$pulse one has 
\end{subequations}
\begin{subequations}
\label{23}
\begin{eqnarray}
\rho _{ee}^{\left( +\right) }\left( \mathbf{x},\mathbf{p},t\right) &=&\rho
_{gg}^{\left( -\right) }\left( \mathbf{x},\mathbf{p}-\hbar \mathbf{k}%
,t\right) ,  \label{23a} \\
\rho _{gg}^{\left( +\right) }\left( \mathbf{x},\mathbf{p},t\right) &=&\rho
_{ee}^{\left( -\right) }\left( \mathbf{x},\mathbf{p}+\hbar \mathbf{k}%
,t\right) ,  \label{23b} \\
\rho _{eg}^{\left( +\right) }\left( \mathbf{x},\mathbf{p},t_{+}\right)
&=&\exp \left[ 2i\left( \mathbf{k}\cdot \mathbf{x}-\delta t-\phi \right) %
\right] \rho _{ge}^{\left( -\right) }\left( \mathbf{x},\mathbf{p},t\right) .
\label{23c}
\end{eqnarray}%
Assume that initially the atoms were in the ground state 
\end{subequations}
\begin{equation}
\rho \left( \mathbf{x},\mathbf{p},0\right) =\left( 
\begin{array}{cc}
0 & 0 \\ 
0 & f\left( \mathbf{x},\mathbf{p}\right)%
\end{array}%
\right) ,  \label{24}
\end{equation}%
where $f\left( \mathbf{x},\mathbf{p}\right) $ is the distribution function
of atoms in the Wigner representation. Applying expressions (\ref{16}, \ref%
{22}, \ref{23}) successively, one obtains for the distribution function of
atoms in the excited state after the third pulse

\begin{subequations}
\label{25}
\begin{eqnarray}
\rho _{ee}^{\left( +\right) }\left( \mathbf{x},\mathbf{p},T_{3}\right) &=&%
\bar{r}\left( \mathbf{x},\mathbf{p}\right) +\tilde{r}\left( \mathbf{x},%
\mathbf{p}\right) ,  \label{25a} \\
\bar{r}\left( \mathbf{x},\mathbf{p}\right) &=&\dfrac{1}{4}\left\{ \dfrac{1}{2%
}f\left[ 
\begin{array}{c}
\mathbf{\xi }\left( \mathbf{\xi }\left( \mathbf{x},\mathbf{p}%
,T_{2},T_{3}\right) ,\mathbf{\pi }\left( \mathbf{x},\mathbf{p}%
,T_{2},T_{3}\right) -\hbar \mathbf{k},0,T_{2}\right) , \\ 
\mathbf{\pi }\left( \mathbf{\xi }\left( \mathbf{x},\mathbf{p}%
,T_{2},T_{3}\right) ,\mathbf{\pi }\left( \mathbf{x},\mathbf{p}%
,T_{2},T_{3}\right) -\hbar \mathbf{k},0,T_{2}\right)%
\end{array}%
\right] \right.  \notag \\
&&+\left. f\left[ 
\begin{array}{c}
\mathbf{\xi }\left( 
\begin{array}{c}
\mathbf{\xi }\left( \mathbf{\xi }\left( \mathbf{x},\mathbf{p}-\hbar \mathbf{k%
},T_{2},T_{3}\right) ,\mathbf{\pi }\left( \mathbf{x},\mathbf{p}-\hbar 
\mathbf{k},T_{2},T_{3}\right) +\hbar \mathbf{k},T_{1},T_{2}\right) , \\ 
\mathbf{\pi }\left( \mathbf{\xi }\left( \mathbf{x},\mathbf{p}-\hbar \mathbf{k%
},T_{2},T_{3}\right) ,\mathbf{\pi }\left( \mathbf{x},\mathbf{p}-\hbar 
\mathbf{k},T_{2},T_{3}\right) +\hbar \mathbf{k},T_{1},T_{2}\right) -\hbar 
\mathbf{k},0,T_{1}%
\end{array}%
\right) , \\ 
\mathbf{\pi }\left( 
\begin{array}{c}
\mathbf{\xi }\left( \mathbf{\xi }\left( \mathbf{x},\mathbf{p}-\hbar \mathbf{k%
},T_{2},T_{3}\right) ,\mathbf{\pi }\left( \mathbf{x},\mathbf{p}-\hbar 
\mathbf{k},T_{2},T_{3}\right) +\hbar \mathbf{k},T_{1},T_{2}\right) , \\ 
\mathbf{\pi }\left( \mathbf{\xi }\left( \mathbf{x},\mathbf{p}-\hbar \mathbf{k%
},T_{2},T_{3}\right) ,\mathbf{\pi }\left( \mathbf{x},\mathbf{p}-\hbar 
\mathbf{k},T_{2},T_{3}\right) +\hbar \mathbf{k},T_{1},T_{2}\right) -\hbar 
\mathbf{k},0,T_{1}%
\end{array}%
\right)%
\end{array}%
\right] \right\} ,  \label{25b} \\
\tilde{r}\left( \mathbf{x},\mathbf{p}\right) &=&-\dfrac{1}{2}\cos \left\{
\left( \mathbf{kx}-\delta _{3}T_{3}-\phi _{3}\right) -2\left[ \mathbf{k}%
\mathbf{\xi }\left( \mathbf{x},\mathbf{p}-\dfrac{\hbar \mathbf{k}}{2}%
,T_{2},T_{3}\right) -\delta _{2}T_{2}-\phi _{2}\right] \right.  \notag \\
&&\left. +\left[ \mathbf{k}\mathbf{\xi }\left( \mathbf{x},\mathbf{p}-\dfrac{%
\hbar \mathbf{k}}{2},T_{1},T_{3}\right) -\delta _{1}T_{1}-\phi _{1}\right]
\right\}  \notag \\
&&f\left\{ 
\begin{array}{c}
\mathbf{\xi }\left[ \mathbf{\xi }\left( \mathbf{x},\mathbf{p}-\dfrac{\hbar 
\mathbf{k}}{2},T_{1},T_{3}\right) ,\mathbf{\pi }\left( \mathbf{x},\mathbf{p}-%
\dfrac{\hbar \mathbf{k}}{2},T_{1},T_{3}\right) -\dfrac{\hbar \mathbf{k}}{2}%
,0,T_{1}\right] , \\ 
\mathbf{\pi }\left[ \mathbf{\xi }\left( \mathbf{x},\mathbf{p}-\dfrac{\hbar 
\mathbf{k}}{2},T_{1},T_{3}\right) ,\mathbf{\pi }\left( \mathbf{x},\mathbf{p}-%
\dfrac{\hbar \mathbf{k}}{2},T_{1},T_{3}\right) -\dfrac{\hbar \mathbf{k}}{2}%
,0,T_{1}\right]%
\end{array}%
\right\} .  \label{25c}
\end{eqnarray}

This expression can be used to calculate any response associated with
excited atoms. The contributions $\bar{r}\left( \mathbf{x},\mathbf{p}\right) 
$ and $\tilde{r}\left( \mathbf{x},\mathbf{p}\right) $ are responsible for
the excitation background and the interference term. We consider the total
probability of the atoms' excitation 
\end{subequations}
\begin{equation}
w\equiv \int d\mathbf{x}d\mathbf{p}\rho _{ee}\left( \mathbf{x},\mathbf{p}%
,T_{3+}\right) =\bar{w}+\tilde{w}.  \label{26}
\end{equation}

Since the phase volume does not change when atoms move along trajectories (%
\ref{18}) as well as with a permanent shift $\pm \hbar \mathbf{k}$ of the
momentum space, the background $\bar{w}=1/2$. In the expression for the
interference term, one can use as integration variables%
\begin{equation}
\left\{ \mathbf{x}^{\prime \prime },\mathbf{p}^{\prime \prime }\right\}
=\left\{ \xi \left( \mathbf{x}^{\prime },\mathbf{p}^{\prime },T_{1},0\right)
,\pi \left( \mathbf{x}^{\prime },\mathbf{p}^{\prime },T_{1},0\right)
\right\} ,  \label{27a}
\end{equation}%
where%
\begin{equation}
\left( 
\begin{array}{c}
\mathbf{x}^{\prime } \\ 
\mathbf{p}^{\prime }%
\end{array}%
\right) =\left( 
\begin{array}{c}
\mathbf{\xi }\left[ \mathbf{\xi }\left( \mathbf{x},\mathbf{p}-\dfrac{\hbar 
\mathbf{k}}{2},T_{1},T_{3}\right) ,\mathbf{\pi }\left( \mathbf{x},\mathbf{p}-%
\dfrac{\hbar \mathbf{k}}{2},T_{1},T_{3}\right) -\dfrac{\hbar \mathbf{k}}{2}%
,0,T_{1}\right] \\ 
\mathbf{\pi }\left[ \mathbf{\xi }\left( \mathbf{x},\mathbf{p}-\dfrac{\hbar 
\mathbf{k}}{2},T_{1},T_{3}\right) ,\mathbf{\pi }\left( \mathbf{x},\mathbf{p}-%
\dfrac{\hbar \mathbf{k}}{2},T_{1},T_{3}\right) -\dfrac{\hbar \mathbf{k}}{2}%
,0,T_{1}\right]%
\end{array}%
\right)  \label{27b}
\end{equation}%
Then one consistently obtains, using the law of multiplication (\ref{18.1}), 
\begin{subequations}
\label{28}
\begin{gather}
\left\{ \mathbf{\xi }\left( \mathbf{x},\mathbf{p}-\dfrac{\hbar \mathbf{k}}{2}%
,T_{1},T_{3}\right) ,\mathbf{\pi }\left( \mathbf{x},\mathbf{p}-\dfrac{\hbar 
\mathbf{k}}{2},T_{1},T_{3}\right) \right\} =\left\{ \mathbf{x}^{\prime
\prime },\mathbf{p}^{\prime \prime }+\dfrac{\hbar \mathbf{k}}{2}\right\} ,
\label{28a} \\
\left\{ \mathbf{x},\mathbf{p}\right\} =\left\{ \mathbf{\xi }\left( \mathbf{x}%
^{\prime \prime },\mathbf{p}^{\prime \prime }+\dfrac{\hbar \mathbf{k}}{2}%
,T_{3},T_{1}\right) ,\mathbf{\pi }\left( \mathbf{x}^{\prime \prime },\mathbf{%
p}^{\prime \prime }+\dfrac{\hbar \mathbf{k}}{2},T_{3},T_{1}\right) +\dfrac{%
\hbar \mathbf{k}}{2}\right\} ,  \label{28b} \\
\mathbf{\xi }\left( \mathbf{x},\mathbf{p}-\dfrac{\hbar \mathbf{k}}{2}%
,T_{i},T_{3}\right) =\mathbf{\xi }\left( \mathbf{x}^{\prime \prime },\mathbf{%
p}^{\prime \prime }+\dfrac{\hbar \mathbf{k}}{2},T_{i},T_{1}\right) ,
\label{28c}
\end{gather}%
yielding this result for the interference term

\end{subequations}
\begin{eqnarray}
\tilde{w} &=&-\dfrac{1}{2}\int d\mathbf{x}d\mathbf{p}f\left[ \mathbf{\xi }%
\left( \mathbf{x},\mathbf{p},0,T_{1}\right) ,\pi \left( \mathbf{x},\mathbf{p}%
,0,T_{1}\right) \right]  \notag \\
&&\times \cos \left[ \phi \left( \mathbf{x},\mathbf{p}\right) -\phi
_{3}+2\phi _{2}-\phi _{1}-\delta _{3}T_{3}+2\delta _{2}T_{2}-\delta _{1}T_{1}%
\right] ,  \label{29}
\end{eqnarray}%
where the AI phase is given by%
\begin{equation}
\phi \left( \mathbf{x},\mathbf{p}\right) =\mathbf{k}\cdot \left[ \mathbf{\xi 
}\left( \mathbf{x},\mathbf{p}+\dfrac{\hbar \mathbf{k}}{2},T_{3},T_{1}\right)
-2\mathbf{\xi }\left( \mathbf{x},\mathbf{p}+\dfrac{\hbar \mathbf{k}}{2}%
,T_{2},T_{1}\right) +\mathbf{x}\right]  \label{30}
\end{equation}%
One replaced $\left\{ \mathbf{x}^{\prime \prime },\mathbf{p}^{\prime \prime
}\right\} \rightarrow \left\{ \mathbf{x},\mathbf{p}\right\} $ in the Eq. (%
\ref{29}).

\section{\label{s4}Approximate approach}

Let us assume that our platform rotation matrix hovers close to identity.%
\begin{equation}
\underline{R}\left( t\right) \approx \underline{I}-\underline{\theta }\left(
t\right) ,  \label{30.1}
\end{equation}%
where $\mathbf{\theta }\left( t\right) $ is the small angle of rotation of
the platform,%
\begin{equation}
\left\vert \mathbf{\theta }\left( t\right) \right\vert \ll 1.  \label{30.2}
\end{equation}%
Let us also assume that the gravity-gradient forces only slightly affect the
atomic trajectory during the interrogation time $T$ 
\begin{equation}
\left\Vert \underline{\mathbf{\gamma }}\left( \mathbf{X}_{c}\right)
\right\Vert T^{2}\ll 1,  \label{31}
\end{equation}%
[where the matrix norm is defined in Eq. (\ref{21})], and 
\begin{equation}
\left\vert \mathbf{\Omega }_{E}\right\vert T\ll 1.  \label{32a}
\end{equation}%
This section focuses solely on terms linear in parameters $\left\{ \mathbf{%
\theta }\left( t\right) ,\underline{\mathbf{\gamma }}\left( \mathbf{X}%
_{c}\right) ,\mathbf{\Omega }_{E}\right\} $. Eqs. (\ref{7c}-\ref{7e}) yield 
\begin{subequations}
\label{33}
\begin{eqnarray}
\underline{x}_{x}\left( \left( t_{2}-t_{1}\right) \right) &\approx &%
\underline{I}+\underline{\mathbf{\gamma }}\left( \mathbf{X}_{c}\right) 
\dfrac{\left( t_{2}-t_{1}\right) ^{2}}{2},  \label{33a} \\
\underline{x}_{v}\left( \left( t_{2}-t_{1}\right) \right) &\approx &%
\underline{I}\left( t_{2}-t_{1}\right) -\underline{\Omega }_{E}\left(
t_{2}-t_{1}\right) ^{2}+\underline{\mathbf{\gamma }}\left( \mathbf{X}%
_{c}\right) \dfrac{\left( t_{2}-t_{1}\right) ^{3}}{6},  \label{33b} \\
\underline{x}_{g} &\approx &\underline{I}\dfrac{\left( t_{2}-t_{1}\right)
^{2}}{2}-\underline{\Omega }_{E}\dfrac{\left( t_{2}-t_{1}\right) ^{3}}{3}+%
\underline{\mathbf{\gamma }}\left( \mathbf{X}_{c}\right) \dfrac{\left(
t_{2}-t_{1}\right) ^{4}}{24}.  \label{33c}
\end{eqnarray}

When one plugs these matrices into Equation (\ref{18a}), one gets the
propagator 
\end{subequations}
\begin{eqnarray}
\mathbf{\xi }\left( \mathbf{x},\mathbf{p},t,t^{\prime }\right) &=&\mathbf{x}+%
\left[ \dfrac{\mathbf{p}}{M}+\mathbf{V}\left( t^{\prime }\right) \right]
\left( t-t^{\prime }\right) +\dfrac{\left( t-t^{\prime }\right) ^{2}}{2}%
\mathbf{g}\left( \mathbf{X}_{c}\right) +\mathbf{X}\left( t^{\prime }\right) -%
\mathbf{X}\left( t\right)  \notag \\
&&+\left[ \underline{\theta }\left( t^{\prime }\right) -\underline{\theta }%
\left( t\right) \right] \left[ \mathbf{x}+\left( t-t^{\prime }\right) \dfrac{%
\mathbf{p}}{M}\right] -\underline{\theta }\left( t\right) \left[ \left(
t-t^{\prime }\right) \mathbf{V}\left( t^{\prime }\right) +\dfrac{\left(
t-t^{\prime }\right) ^{2}}{2}\mathbf{g}\left( \mathbf{X}_{c}\right) +\mathbf{%
X}\left( t^{\prime }\right) -\mathbf{X}\left( t\right) \right]  \notag \\
&&-\underline{\Omega }_{E}\left\{ \left( t-t^{\prime }\right) \mathbf{x}%
+\left( t-t^{\prime }\right) ^{2}\left[ \dfrac{\mathbf{p}}{M}+\mathbf{V}%
\left( t^{\prime }\right) \right] +\dfrac{\left( t-t^{\prime }\right) ^{3}}{3%
}\mathbf{g}\left( \mathbf{X}_{c}\right) \right\}  \notag \\
&&+\underline{\mathbf{\gamma }}\left( \mathbf{X}_{c}\right) \left\{ \dfrac{%
\left( t-t^{\prime }\right) ^{2}}{2}\mathbf{x}+\dfrac{\left( t-t^{\prime
}\right) ^{3}}{6}\left[ \dfrac{\mathbf{p}}{M}+\mathbf{V}\left( t^{\prime
}\right) \right] +\dfrac{\left( t-t^{\prime }\right) ^{4}}{24}\mathbf{g}%
\left( \mathbf{X}_{c}\right) +\dfrac{\left( t-t^{\prime }\right) ^{2}}{2}%
\left[ \mathbf{X}\left( t^{\prime }\right) -\mathbf{X}_{c}\right] \right\}
\label{34}
\end{eqnarray}

Let's look at when the field pulses occur at the times: $\left\{
T_{1},T_{2},T_{3}\right\} =\left\{ 0,T,2T\right\} $. Then the AI phase (\ref%
{30}) is given by 
\begin{subequations}
\label{35}
\begin{eqnarray}
\phi \left( \mathbf{x},\mathbf{p}\right) &=&\phi _{A}+\phi _{\underline{%
\Omega }}+\phi _{\underline{\mathbf{\gamma }}}+\phi _{x}+\phi _{X}+\phi
_{\theta },  \label{35a} \\
\phi _{A} &=&T^{2}\mathbf{k\cdot }\left[ \mathbf{g}\left( \mathbf{X}%
_{c}\right) -\dfrac{\mathbf{X}\left( 0\right) -2\mathbf{X}\left( T\right) +%
\mathbf{X}\left( 2T\right) }{T^{2}}\right] ,  \label{35b} \\
\phi _{\Omega } &=&2T^{2}\left[ \mathbf{\Omega }_{E}+\dfrac{\mathbf{\theta }%
\left( 2T\right) -\mathbf{\theta }\left( T\right) }{T}\right] \cdot \left\{ 
\mathbf{k}\times \left[ \dfrac{\mathbf{p}}{M}+\mathbf{V}\left( 0\right) +%
\mathbf{g}\left( \mathbf{X}_{c}\right) T\right] \right\} ,  \label{35c} \\
\phi _{\mathbf{\gamma }} &=&\mathbf{k\cdot }\underline{\mathbf{\gamma }}%
\left( \mathbf{X}_{c}\right) T^{2}\left\{ \mathbf{x}+\mathbf{X}\left(
0\right) -\mathbf{X}_{c}+T\left[ \left( \dfrac{\mathbf{p}}{M}+\dfrac{\hbar 
\mathbf{k}}{2M}\right) +\mathbf{V}\left( 0\right) \right] +\dfrac{7}{12}T^{2}%
\mathbf{g}\left( \mathbf{X}_{c}\right) \right\} ,  \label{35d} \\
\phi _{x} &=&\left[ \mathbf{\theta }\left( 0\right) -2\mathbf{\theta }\left(
T\right) +\mathbf{\theta }\left( 2T\right) \right] \cdot \left( \mathbf{k}%
\times \mathbf{x}\right) ,  \label{35e} \\
\phi _{X} &=&-\mathbf{\theta }\left( 2T\right) \cdot \left\{ \mathbf{k}%
\times \left[ \mathbf{X}\left( 2T\right) -\mathbf{X}\left( 0\right) \right]
\right\} +2\mathbf{\theta }\left( T\right) \cdot \left\{ \mathbf{k}\times %
\left[ \mathbf{X}\left( T\right) -\mathbf{X}\left( 0\right) \right] \right\}
\label{35f} \\
\phi _{\theta } &=&T^{2}\mathbf{\theta }\left( T\right) \cdot \left[ \mathbf{%
k}\times \mathbf{g}\left( \mathbf{X}_{c}\right) \right] .  \label{35g}
\end{eqnarray}%
One can get these approximate results by doing the math in the inertial
frame.

\section{\label{s5}Discussion}

Looking beyond the boundaries of atomic interferometry, the presented here
consideration yields navigation equations (\ref{6.1}). These equations are a
generalization of the navigation equations (4.77) in the textbook \cite{c37}%
, consisting first in that one takes into account the gravity-gradient
forces, and second in that one considers the motion and rotation of the
platform not relative to the inertial frame, but relative to the
non-inertial frame RGS. The advantage of this choice is that only in this
frame the gravitational potential (\ref{5}) is independent of time and
therefore the equations of motion contain only time-independent coefficients
and time-independent right-hand-sides.

In this paper we have derived all expressions to calculate the phase of the
AI by the path integral approach. Namely, the Lagrangian (\ref{6a}) and the
exact trajectory of the atom in the Lagrangian variables $\left\{ \mathbf{x},%
\mathbf{v}\right\} $ (\ref{2}, \ref{4}, \ref{7}) are obtained. Following the
technique described many times in the papers \cite%
{c10,c12,c12.1,c13,c14,c15,c16,c16.1,c17,c18,c19,c20,c21,c22}, one could, by
calculating the action for 4 trajectories in the recoil diagram of the
Mach-Zehnder AI, obtain the desired phase. The difficulty I encountered and
could not overcome was that the integrands in the phase factors are
quadratic in the variables $\left\{ \mathbf{x},\mathbf{v}\right\} $ and, in
addition, they still need to be integrated over time, substituting the exact
trajectory (\ref{2}, \ref{4}, \ref{7}) \ Moreover, one should perform this
integration for an arbitrarily time-dependent platform position and
orientation. In contrast, the phase factor in the expression for the
interference part of the density matrix in the Wigner representation (\ref%
{25c}) is linear in the atomic trajectory in platform frame and does not
need to be integrated.

Formula (\ref{30}) is an exact analytical expression for the phase of the
AI. In it, the phase is expressed in terms of the positions of the atom ($%
\xi -$propagators) at the moments of action of the Raman field pulses,
taking into account the quantum jump of the atomic momentum$\dfrac{\hbar 
\mathbf{k}}{2}$ arising from the interaction with this field. Without taking
into account the quantization of the motion of the atomic center-of-mass, a
similar expression for the phase was used in the theory of the atomic
navigator \cite{c29,c30}. One obtains Eq. (\ref{18}) for the propagators in
the phase space for an arbitrary motion and rotation of the platform, and
the matrices (\ref{7c} - \ref{7e}) were calculated \cite{c31} for an
arbitrary relationship between, on the one hand, the interrogation time $T$
(with which the time $t-t^{\prime }$ in Eq. (\ref{7a}) is comparable), and,
on the other hand, the gravity-gradient tensor \underline{$\mathbf{\gamma }$}%
$\left( \mathbf{X}\right) $ and the permanent rotation rate of the gravity
source \textbf{$\Omega $}$_{E}$. Let us recall, that when one did the math
using path integrals \cite%
{c10,c12,c12.1,c13,c14,c15,c16,c16.1,c17,c18,c19,c20,c21,c22}, one treated
the platform as performing uniformly accelerated motion and rotating at a
small, permanent rate $\mathbf{\Omega }$ throughout time $T$. One obtains
the exact result (\ref{30}) because (i) the quantum equation for the density
matrix in the Wigner representation [for the gravitational potential (\ref{5}%
)] coincides with the classical Liouville equation, which has simple
well-known solution (\ref{16}), and (ii) one used the exact trajectory of an
atom (\ref{7}) in the field of a RGS \cite{c31}. Path integrals can be
calculated numerically for any specific platform motion and rotation,
yielding an AI phase that matches our analytical result.

We derived the exact phase (\ref{30}) of the integrand in the total
excitation probability's interfering term for the atomic cloud. Calculation
for arbitrary Raman pulse action times $T_{i}$. The integrand oscillates
rapidly as a function of atomic momentum, with period $\sim M/kT$, washing
out interference when averaging over momenta. In the language of Wigner
representation, the interferometer is closed at that critical value of $T_{i}
$, the precise point where oscillations cease to exist. If, despite
rotation, one can find these values of $T_{i}$, then one observes
macroscopic interference.

In the exact expression for the phase AI all the factors included in the
consideration act simultaneously and interdependently. One is not able to
separate one factor from another. Such a possibility appears when passing to
the approximate expression (\ref{35}). Regarding these formulas one may note
that we have not encountered the terms $\phi _{x},$ $\phi _{X}$ and $\phi
_{\theta }$ in the literature, and they are apparently new. The term $\phi
_{x}$ arises when the platform rotates with a time-varying rate \textbf{$%
\Omega $}$\left( t\right) $. The term $\phi _{X}$ is a consequence of a
combination of rotational and translational motion of the platform, while
the term $\phi _{\theta }$ appears because the RGS gravitational field
rotates in the platform's frame. Similar to the main term $\phi _{A}$, both $%
\phi _{X}$ and $\phi _{\theta }$ remain unaffected by an atom's initial
position in the $\left\{ \mathbf{x},\mathbf{p}\right\} $ phase space -
rendering them invisible to atomic gravity-gradiometers and gyroscopes that
rely on the differential technique. One may also note that Eqs. (\ref{35})
does not coincide with the result obtained in \cite{c27}, since the latter
lacks terms containing the rotation rate of the gravity source \textbf{$%
\Omega $}$_{E}$. On the other hand, if we restrict ourselves to terms up to $%
T^{4}$, then for a stationary and non-rotating platform 
\end{subequations}
\begin{subequations}
\label{36}
\begin{eqnarray}
\mathbf{X}\left( t\right) &=&\mathbf{X}_{c},  \label{36a} \\
\mathbf{\theta }\left( t\right) &=&0,  \label{36b}
\end{eqnarray}%
one will arrive at the result obtained in \cite{c6,c15}, in which one
assumes a spherical shape of the Earth and neglects the contribution of
anomaly gravity. The contribution of anomaly gravity becomes dominant only
for higher-order gravity-gradient tensors \cite{c36}.

Various terms in Eqs. (\ref{35}) allow one to use AIs as atomic gravimeters,
accelerometers, gyroscopes and navigators. Only one term corresponds to a
given AI application, while the others lead to systematic or statistical
measurement errors. Studying them and developing methods for their
elimination allows one to predict the expected measurement accuracy. Term (%
\ref{35b}) $\phi _{A}$ one uses to measure the gravitational field $\mathbf{g%
}\left( \mathbf{X}\right) $ \cite{c10} and to test Einstein's equivalence
principle \cite{c38}. Even in the ideal case, when the conditions (\ref{36})
are met, the atoms are launched from the origin with zero initial momentum
and 
\end{subequations}
\begin{equation}
\mathbf{k}\parallel \mathbf{g}\left( \mathbf{X}_{c}\right) ,  \label{36.1}
\end{equation}%
gravity-gradient term%
\begin{equation}
\phi _{\mathbf{\gamma }}=T^{3}\mathbf{k\cdot }\underline{\mathbf{\gamma }}%
\left( \mathbf{X}_{c}\right) \dfrac{\hbar \mathbf{k}}{2M}+\dfrac{7}{12}T^{4}%
\mathbf{k\cdot }\underline{\mathbf{\gamma }}\left( \mathbf{X}_{c}\right) 
\mathbf{g}\left( \mathbf{X}_{c}\right)  \label{37}
\end{equation}%
leads to a systematic measurement error of $\mathbf{g}\left( \mathbf{X}%
\right) $ \cite{c39}. The first term here is the quantum correction the
phase of the AI. It was observed in \cite{c40}. Only this term confirms that
the AI is a quantum sensor. One can also observe the quantum phase in a
uniform gravitational field [when $\underline{\mathbf{\gamma }}\left( 
\mathbf{X}_{c}\right) =0$] using the technique of sequential momentum
transfer, when it increases by $\left[ \left\Vert \underline{\mathbf{\gamma }%
}\left( \mathbf{X}_{c}\right) \right\Vert T^{2}\right] ^{-1}$ times
achieving its maximum value \cite{c41}. The second term in Eq. (\ref{37})
arises as a result of the gravitational field change along the atomic
trajectory during the interrogation time $T$ \cite{c42}.

Consider now a moving but not rotating platform. If the gravitational field
is known, one measures the second term in square brackets in Eq. (\ref{35b}%
). For a small interrogation time, this term is $-$\textbf{$A$}$\left(
0\right) $, i.e. the AI is an accelerometer. Then, using the standard
navigation procedure \cite{c37}, substituting \textbf{$A$}$\left( t\right) $
into the navigation equations and solving them numerically, one determines
the position of the platform. From the point of view of atomic
interferometry, this procedure looks rather strange. Why follow it when one
can immediately determine the position of the platform $\mathbf{X}\left(
2T\right) $ from the phase, and preceding platform positions $\mathbf{X}%
\left( T\right) $ and $\mathbf{X}\left( 0\right) $? By exploiting the AIs at
multiple times $nT$ with zero dead time (ZDT) \cite{c42.1}, one determines
the position of the platform at times $\left( n+2\right) T$. This ZDT
approach to navigation was proposed in patent \cite{c29} and demonstrated in 
\cite{c43}. The most important thing about ZDT \cite{c31} is that one does
not need navigation equations at all. This means one can make the
interrogation time $T$ as long as possible to get the best navigation
accuracy. In conventional navigation, one cannot do this because $T$ is
restricted by the time step of the navigation equations' numerical solution.

One associates the term $\mathbf{X}\left( t\right) $. with vibrational
noise, which affects the operation of both the gravimeter and the
accelerometer. If, however, the ZDT is used for navigation, then the
navigation noise is not important, since in this approach one immediately,
by measuring the phase of the AI, reaches the final navigation goal, the
platform position $\mathbf{X}\left( 2T\right) $, including the instantaneous
value of the vibration noise at $t=2T$. The authors considered vibration
noise in projects \cite%
{c1.2,c1.3,c1.5,c1.5.2,c1.5.3,c1.5.7,c1.5.9,c1.5.4,c1.5.8}. Here one can use
seismometer data \cite{c1.3}.

One can get rid of vibrational noise by using the differential technique 
\cite{c44}. The phase difference between identical AIs launched at $\mathbf{x%
}_{1}$ and $\mathbf{x}_{2}$,%
\begin{equation}
\Delta \phi =T^{2}\mathbf{k}\underline{\mathbf{\gamma }}\left( \mathbf{X}%
_{c}\right) \Delta \mathbf{x},  \label{38}
\end{equation}%
where $\Delta \mathbf{x}=\mathbf{x}_{1}-\mathbf{x}_{2}$, measures the
gravity gradient tensor \underline{$\mathbf{\gamma }$} \ The differential
technique is widely used in testing Einstein's equivalence principle \cite%
{c38}, measuring Newton's gravitational constant \cite{c45}, one proposes it
for observing gravitational waves \cite{c46} and for other effects.

The rotation of the platform leads to systematic and statistical errors in
atomic gravimeters, accelerometers and gravity-gradiometers. Even for
differential technique term (\ref{35e}) does not disappear and one must add
term%
\begin{equation}
\Delta \phi _{x}=\left[ \mathbf{\theta }\left( 0\right) -2\mathbf{\theta }%
\left( T\right) +\mathbf{\theta }\left( 2T\right) \right] \mathbf{k}\times
\Delta \mathbf{x}  \label{39}
\end{equation}%
to the gradiometer phase (\ref{38}). Rotational noise can be reduced by
placing the AI on a heavy stabilization stand. This possibility, however,
vanishes for transportable AIs and those operating in the microgravity
environment. Then one can use a gimbal with rms $\theta \sim 10^{-5}$rad as
a platform. To ignore the addition (\ref{39}) it must be smaller than the
phase measurement error $\phi _{err}$. The minimum phase difference
measurement error $\phi _{err}\sim 2\times 10^{-5}$rad was achieved in Ref. 
\cite{c47}. Then one obtains for a small angle $\alpha $ between vectors $%
\mathbf{k}$ and $\Delta \mathbf{x}$%
\begin{equation}
\alpha \lesssim \phi _{err}/k\theta \Delta x.  \label{40}
\end{equation}%
For typical values of $k\sim 10^{7}$m$^{\text{-1}},$ $\Delta x\sim 10$m one
finds that the parallelism of vectors $\mathbf{k}$ and $\Delta \mathbf{x}$
should be maintained with an accuracy of $\alpha \lesssim 2\times 10^{-8}$%
rad.

One can use the AI as a gyroscope \cite{c48}. In atomic gyroscopes one split
atomic states using microfabricated structures \cite{c49} or Raman fields 
\cite{c50}. One considers this second case here. If one abandons the
conditions (\ref{36}, \ref{36.1}), then the term $\phi _{\Omega }$ in (\ref%
{35c}) corresponds to a gyroscope. Earlier we considered the differential
approach in coordinate space, now we will consider it in momentum space: if
in two identical AIs the atomic clouds are launched from the same point $%
\mathbf{x}$, but with two momenta $\mathbf{p}_{1}$ and $\mathbf{p}_{2}$,
then the phase difference of these AIs is%
\begin{equation}
\Delta \phi =2T^{2}\left[ \mathbf{\Omega }_{E}+\dfrac{\mathbf{\theta }\left(
2T\right) -\mathbf{\theta }\left( T\right) }{T}\right] \cdot \left( \mathbf{k%
}\times \dfrac{\Delta \mathbf{p}}{M}\right) +\mathbf{k}\underline{\mathbf{%
\gamma }}\left( \mathbf{X}_{c}\right) T^{3}\dfrac{\Delta \mathbf{p}}{M},
\label{41}
\end{equation}%
where $\Delta \mathbf{p}=\mathbf{p}_{1}-\mathbf{p}_{2}$. This equation's
second term causes the gyroscope systematic error. The relative weight of
this term is $\dfrac{\left\Vert \underline{\mathbf{\gamma }}\left( \mathbf{X}%
_{c}\right) \right\Vert T}{\left\vert \mathbf{\Omega }_{E}\right\vert }\sim
4\times 10^{-2}$ (for $T\sim 1$s, in the case of the Earth's gravitational
field, when $\left\Vert \underline{\mathbf{\gamma }}\left( \mathbf{X}%
_{c}\right) \right\Vert \sim 3\times 10^{3}$E). One can reduce this
contribution by choosing the directions of the vectors $\mathbf{k}$ and $%
\Delta \mathbf{p}$ along different axes of the North-East-Down frame, when
the systematic error is proportional to the off-diagonal element of the
tensor \underline{$\mathbf{\gamma }$}$\left( \mathbf{X}_{c}\right) $, and it
is two orders of magnitude smaller than the diagonal elements either due to
the small value of the Geoid eccentricity or due to the small contribution
of the Earth's anomaly gravity to the first-order gravity-gradient tensor 
\cite{c51}. The systematic error associated with the centrifugal term in Eq.
(\ref{6d}) will be even smaller, $\sim \left\vert \mathbf{\Omega }%
_{E}\right\vert T\sim 7\times 10^{-5}.$

If the platform is protected from rotation, then one measures the rotation
rate of the gravity source \textbf{$\Omega $}$_{E}$. Another possibility for
measuring \textbf{$\Omega $}$_{E}$ is the $T^{3}-$term in Eq. (\ref{35c}).
To select this term, one uses the double-loop technique \cite{c52}. In this
context, a specific issue has been identified \cite{c6}: the two-loop $\pi
/2-\pi -\pi -\pi /2$ AI lacks $T^{2}-$terms, but it aligns with the timing
of Raman pulses for a stimulated echo ($\pi /2-\pi /2-\pi /2-\pi /2$ AI), in
which the phase contains a significant term $\sim \mathbf{k}\cdot \mathbf{g}%
\left( \mathbf{X}_{c}\right) T^{2}$. To kill this term, one has proposed 
\cite{c6} and implemented \cite{c53} the method of adjustable momentum
transfer. Here one slightly modifies the wave vectors, separating in time
the useful gyroscope signal from the parasitic stimulated echo. In an
alternative method \cite{c54} one reduces the delay between the 1st and 2nd
and increases the delay between the 3rd and 4th pulses, which also separates
the useful and parasitic signals, but on the other hand leads to a small $%
T^{2}-$term. This method achieves $1\times 10^{-10}$rad/s sensitivity \cite%
{c55} for an atomic gyroscope.

If the rotation rate of the gravity source is known, one can use an atomic
gyroscope (\ref{41}) for navigation. For a sufficiently small interrogation
time $T$, the second term in square brackets is the rotation rate of the
platform relative to the RGS, \textbf{$\Omega $}$\left( 0\right) $. From the
phase difference (\ref{41}) one determines the projection \textbf{$\Omega $}$%
\left( 0\right) $ on the interrogating axis $\mathbf{k}\times \Delta \mathbf{%
p}$. To determine all projections \textbf{$\Omega $}$\left( 0\right) $ three
pairs of AIs are needed, each of which has a common effective wave vector $%
\mathbf{k}$ and a common launch point $\mathbf{x}$, i.e. a total of six AIs
are needed. Such an AI system was created in the article \cite{c57}.
However, in it the atomic clouds were launched from different points and
therefore the differences (\ref{38}, \ref{39}) could lead to errors in
measuring \textbf{$\Omega $}$\left( 0\right) $. The measured platform
rotation rate is substituted into the navigation equations, solving which
the orientation and then the position of the platform are determined. The
situation here is as strange as in the case of the atomic accelerometer
above. Why use navigation equations when the phase difference (\ref{41}) and 
$\mathbf{\theta }\left( T\right) $ directly determine $\mathbf{\theta }%
\left( 2T\right) $, including both smooth and noise components? One can put
this alternative into action using the ZDT process \cite{c29,c30,c43}.

We found consideration of rotational noise only in projects \cite%
{c1.5.4,c1.5.8}. In them, the authors considered only the effect of Coriolis
acceleration $-2$\textbf{$\Omega $}$\left( t\right) \times \dfrac{\Delta 
\mathbf{p}}{M}$. One sees from Eq. (\ref{41}) that the problem is reduced to
this acceleration only with a smooth change in orientation over the
interrogation time $T$. To be smooth during this time, the rotational noise
bandwidth $f$ should be sufficiently small, $f\ll T^{-1}$. For example, the
ultimate goal of the project \cite{c1.5.8} is the interrogation time $T=5$s,
then their consideration is valid only with a noise bandwidth $f\ll 2\pi
\times 30$mHz. Otherwise, one should use Eq. (\ref{41}). In the opposite
limit, $f\gg T^{-1}$ the orientations at moments $T$ and $2T$ are not
correlated and if the standard deviation of orientation is $\delta \theta
\sim 10^{-5}$rad, $k\sim 10^{7}$m$^{\text{-1}},\dfrac{\Delta p}{M}\sim 1$%
m/s, and $\mathbf{k}\perp \Delta \mathbf{p}$, then the standard deviation of
the phase difference will be 
\begin{equation}
\delta \Delta \phi =2\sqrt{\dfrac{2}{3}}\delta \theta Tk\dfrac{\Delta p}{M}%
\sim 800\text{rad}.  \label{41.1}
\end{equation}

One can see that only the atomic ZDT navigator \cite{c29} (or its classical
version, the Local Position System \cite{c31}) isn't affected by vibrational
or rotational noises. One finds in papers \cite{c29,c30} a simulation of the
ZDT navigator, in which one used the exact expression for the AI phase
without taking into account the recoil effect. The authors considered two
situations, when there is an accurate map of the RGS gravity field, and when
one uses a system of atomic gravity-gradiometers to measure the
gravity-gradient tensor \underline{$\mathbf{\gamma }$}$\left[ \mathbf{X}%
\left( 2T\right) \right] $ and calculate the increment of the gravity field
along the platform trajectory. In both situations, the obtained or measured
values of $\mathbf{g}\left[ \mathbf{X}\left( 0\right) \right] $ and 
\underline{$\mathbf{\gamma }$}$\left[ \mathbf{X}\left( 0\right) \right] $
were substituted into the exact expressions for the phases of the 3 pairs of
AIs. Instead of the usual use \cite{c57} of the 3 phase differences of the
AIs to measure the platform rotation rate \textbf{$\Omega $}$\left( 0\right) 
$, i.e. a quantity that is not needed in the ZDT process, the authors used
the phase differences to determine the orientation of the platform 
\underline{$R$}$\left( 2T\right) $. This was possible because the phase
difference of two AIs with identical effective wave vectors, in which atomic
clouds are launched from one point, depends only on the orientations of the
platform $\left\{ \underline{R}\left( 0\right) ,\underline{R}\left( T\right)
,\underline{R}\left( 2T\right) \right\} $. One sees from Eq. (\ref{41}) that
when rotating at small angles one can finds out the rotation angle during
the action of the AI%
\begin{equation}
\mathbf{\sigma }=\mathbf{\theta }\left( 2T\right) -\mathbf{\theta }\left(
0\right)  \label{42}
\end{equation}%
if the preceding rotation angle $\mathbf{\theta }\left( T\right) -\mathbf{%
\theta }\left( 0\right) $ and the 3 phase differences of the AIs are known.
In the paper \cite{c30} the authors proposed to use $\mathbf{\sigma }$ as an
independent variable also in the case of rotation by a finite angle and to
represent the rotation matrix \underline{$r$}$\equiv $\underline{$R$}$\left(
2T\right) $\underline{$R$}$^{T}\left( 0\right) $ as a series in powers of $%
\mathbf{\sigma }$. One can find in Ref. \cite{c30} the recurrence relations
for the coefficients of this series.

The exact result obtained in this paper (\ref{30}) one can use to generalize
the results of \cite{c29,c30}.

Another application of the result (\ref{30}) is the simulation of the
influence of vibrational and rotational noise on the operation of atomic
gravimeters, gravity-gradiometers and gyroscopes. Examples of such
simulation one can find in the articles \cite{c29,c30,c31}. Instead of
simulation, one can use vibration seismometers \cite{c1.3} and rotational
seismographs \cite{c58}. The data from these sensors can be substituted into
Eq. (\ref{30}) in order to extract the gravitational and inertial parameters
of the RGS from the comparison of the measured phase and its exact value.


\begin{thebibliography}{99}
\bibitem{c1} B. Y. Dubetski\"{\i}, A P. Kazantsev, V. P. Chebotayev, and V.
P. Yakovlev, Interference of atoms and formation of atomic spatial arrays in
light fields, Pis'ma Zh. Eksp. Teor. Fiz. \textbf{39,} 531 (1984) [JETP
Lett. \textbf{39}, 649 (1984)].

\bibitem{c1.0} \textit{Atom Interferometry}, edited by P. R. Berman
(Academic, Cambridge, 1997)

\bibitem{c1.1} \textit{Atom Interferometry}, Proceedings of the
International School of Physics \textquotedblleft Enrico
Fermi,\textquotedblright\ Course CLXXXVIII, Varenna, 2013, edited by G. M.
Tino and M. Kasevich (IOS, Amsterdam, 2014).

\bibitem{c1.1.6} R. Geiger, A. Landragin, S. Merlet, F. P. D. Santos,
High-accuracy inertial measurements with cold-atom sensors, AVS Quantum Sci. 
\textbf{2}, 024702 \href{https://doi.org/10.1116/5.0009093}{(2020).}

\bibitem{c1.1.7} Q. Zhang, Y. Wang, C. Zhu, Y. Wang, X. Zhang, K. Gao, and
W. Zhang, Precision measurements with cold atoms and trapped ions, Chin.
Phys. B \textbf{29}, 093203 \href{https://doi.org/10.1088/1674-1056/aba9c6}{%
(2020).}%
%TCIMACRO{\TeXButton{comment}{\begin{comment}}}%
%BeginExpansion
\begin{comment}%
%EndExpansion
arXiv:2007.09064 [physics.atom-ph] 
%TCIMACRO{\TeXButton{endcomment}{\end{comment}}}%
%BeginExpansion
\end{comment}%
%EndExpansion

\bibitem{c1.1.5} G. M. Tino, Testing gravity with cold atom interferometry:
Results and prospects, in \textit{Focus on Quantum Sensors for New-Physics
Discoveries,} edited by M. Safronova and D. Budker, special issue of Quantum
Sci. Technol. \textbf{6}, 024014 \href{https://doi.org/10.1088/2058-9565/abd83e}%
{(2021).}

\bibitem{c1.1.4} A. Bertoldi, P. Bouyer, and B. Canuel, Quantum sensors with
matter waves for GW observation, in \textit{Handbook of Gravitational Wave
Astronomy}, edited by C. Bambi, S. Katsanevas, and K. D. Kokkotas (Springer,
Singapore, \href{https://doi.org/10.1007/978-981-15-4702-7_5-1}{2021}).

\bibitem{c1.1.3} C. Ufrecht, A. Roura, and W. P. Schleich, Bose-Einstein
condensates in microgravity and fundamental tests of gravity, \href{https://doi.org/10.48550/arXiv.2107.03709}%
{arXiv:2107.03709v1 [quant-ph].}

\bibitem{c1.1.2} A. Belenchia et al., Quantum physics in space, Physics
Reports \textbf{951}, 1 \href{https://doi.org/10.1016/j.physrep.2021.11.004}{%
(2022).}

\bibitem{c1.1.1} I. Alonso et al., Cold atoms in space: Community workshop
summary and proposed road-map. EPJ Quantum Technol. \textbf{9,} 30 \href{https://doi.org/10.1140/epjqt/s40507-022-00147-w}%
{(2022).}

\bibitem{c1.1.0} S. Abend et al., Technology roadmap for cold-atoms based
quantum inertial sensor in space, AVS Quantum Sci. \textbf{5,} 019201 \href{https://doi.org/10.1116/5.0098119}%
{(2023).}

\bibitem{c1.1.8} J. R. Williams et al., Interferometry of Atomic Matter
Waves in the Cold Atom Lab onboard the International Space Station, \href{https://doi.org/10.48550/arXiv.2402.14685}%
{arXiv:2402.14685 [physics.atom-ph]}

\bibitem{c1.2} B. Canuel et al., ELGAR---A European laboratory for
gravitation and atom-interferometric research, Class. Quantum Grav. \textbf{%
37} 225017 \href{https://doi.org/10.1088/1361-6382/aba80e}{(2020).}

\bibitem{c1.3} Ming-Sheng Zhan et al., ZAIGA: Zhaoshan long-baseline atom
interferometer gravitation antenna, Int. J. Mod. Phys. D \textbf{29}, 194005 
\href{https://doi.org/10.1142/S0218271819400054}{ (2020).}

\bibitem{c1.4} L. Badurina et al., AION: An atom interferometer observatory
and network, J. Cosmol. Astropart. Phys. \href{https://doi.org/10.1088/1475-7516/2020/05/011}%
{JCAP05(2020)011}

\bibitem{c1.5} B. Battelier et al., Exploring the foundations of the
universe with space tests of the equivalence principle, Exp. Astron. \textbf{%
51}, 1695 \href{https://doi.org/10.1007/s10686-021-09718-8}{(2021).}

\bibitem{c1.5.1} Y. A. El-Neaj et al., AEDGE: Atomic experiment for dark
matter and gravity exploration in space, EPJ Quantum Technol. \textbf{7}, 6 
\href{https://doi.org/10.1140/epjqt/s40507-020-0080-0}{(2020).}

\bibitem{c1.5.2} G. M. Tino et al., SAGE: A proposal for a space atomic
gravity explorer, Eur. Phys. J. D.\textbf{\ 73}, 228 \href{73, 228 (2019).}{%
(2019).}

\bibitem{c1.5.3} M. Abe et al., Matter-wave Atomic Gradiometer
Interferometric Sensor (MAGIS-100), Quantum Sci. Technol. \textbf{6,} 044003 
\href{https://doi.org/10.1088/2058-9565/abf719}{(2021).}

\bibitem{c1.5.6} Rainer Kaltenbaek et al., Quantum technologies in space,
Exp. Astron. \textbf{51}, 1677 \href{https://doi.org/10.1007/s10686-021-09731-x}%
{(2021).}

\bibitem{c1.5.7} N. Zahzam et al, Hybrid electrostatic-atomic accelerometer
for future space gravity missions, Remote Sens. \textbf{14}, 3273 \href{https://doi.org/10.3390/rs14143273}%
{(2022)}.

\bibitem{c1.5.5} N. Gaaloul et al.,Space Time Explorer and QUantum
Equivalence principle Space Test: The 2022 medium-class mission concept,
arXiv:\href{https://doi.org/10.48550/arXiv.2211.15412}{2211.15412
[physics.space-ph].}

\bibitem{c1.5.9} J. Carlton, C. McCabe, From RATs to riches: mitigating
anthropogenic and synanthropic noise in atom interferometer searches for
ultra-light dark matter, arXiv:\href{https://doi.org/10.48550/arXiv.2308.10731}%
{2308.10731 [astro-ph.CO]}

\bibitem{c1.5.4} Q. Beaufils et al, Rotation related systematic effects in a
cold atom interferometer onboard a Nadir pointing satellite, npj Micrograv.
9, 53 \href{https://doi.org/10.1038/s41526-023-00297-w}{(2023).}%
%TCIMACRO{\TeXButton{comment}{\begin{comment}}}%
%BeginExpansion
\begin{comment}%
%EndExpansion
\U{440}\U{430}\U{441}\U{441}\U{43c}\U{430}\U{442}\U{440}\U{438}\U{432}\U{430}%
\U{43b} \U{432}\U{440}\U{430}\U{449}\U{430}\U{442}\U{435}\U{43b}\U{44c}%
\U{43d}\U{44b}\U{439} \U{448}\U{443}\U{43c} \U{442}\U{43e}\U{43b}\U{44c}%
\U{43a}\U{43e} \U{434}\U{43b}\U{44f} \U{43c}\U{430}\U{43b}\U{44b}\U{445} 
\U{422} 
%TCIMACRO{\TeXButton{endcomment}{\end{comment}}}%
%BeginExpansion
\end{comment}%
%EndExpansion

\bibitem{c1.5.8} A. HosseiniArania et al, Advances in Atom Interferometry
and their Impacts on the Performance of Quantum Accelerometers On-board
Future Satellite Gravity Missions, \hypertarget{%
https://doi.org/10.48550/arXiv.2404.10471}{arXiv:\href{https://doi.org/10.48550/arXiv.2404.10471}%
{2404.10471 [physics.ins-det]}}%
%TCIMACRO{\TeXButton{comment}{\begin{comment}}}%
%BeginExpansion
\begin{comment}%
%EndExpansion
\U{440}\U{430}\U{441}\U{441}\U{43c}\U{430}\U{442}\U{440}\U{438}\U{432}\U{430}%
\U{43b} \U{432}\U{440}\U{430}\U{449}\U{430}\U{442}\U{435}\U{43b}\U{44c}%
\U{43d}\U{44b}\U{439} \U{448}\U{443}\U{43c} \U{442}\U{43e}\U{43b}\U{44c}%
\U{43a}\U{43e} \U{434}\U{43b}\U{44f} \U{43c}\U{430}\U{43b}\U{44b}\U{445} 
\U{422}=5s 
%TCIMACRO{\TeXButton{endcomment}{\end{comment}}}%
%BeginExpansion
\end{comment}%
%EndExpansion

\bibitem{c1.5.10} K. Bongs et al., AION-10: Technical Design Report for a
10m Atom Interferometer in Oxford, \href{https://doi.org/10.48550/arXiv.2508.03491}%
{arXiv:2508.03491 [physics.atom-ph]}.

\bibitem{c2} A. P. Kol'chenko, S. G. Rautian, and R. I. Sokolovskii,
Interaction of an atom with a strong electromagnetic field with the recoil
effect taken into consideration, Zh. Eksp. Teor. Fiz. \textbf{55}, 1864
(1968) [Sov. Phys. JETP \textbf{28}, 986 (1969)] .

\bibitem{c3} L.D.Landau, E.M. Lifshitz, Statistical Physics, Part 1;
Pergamon Press: New York, NY, USA, 1980; p. 20.

\bibitem{c4} Y.L. Klimontovich, Kinetic Theory of Nonideal Gases amd
Nonideal Plasmas; Pergamon Press: New York, NY, USA, 1982; Chapter 12.

\bibitem{c5} P.R. Berman, V.S. Malinovsky, Prinsiples of Laser Spectroscopy
and Quantum Optics; Prinston University Press: Prinston, NJ, USA, 2011;
Section 18.5.

\bibitem{c6} B. Dubetsky, M.A. Kasevich, Atom interferometer as a selective
sensor of rotation or gravity. Phys. Rev. A \textbf{74}, 023615 \href{https://doi.org/10.1103/PhysRevA.74.023615}%
{(2006)}.

\bibitem{c7} B. Dubetsky, S. B. Libby and P. Berman, Atom Interferometry in
the Presence of an External Test Mass\_Atoms \textbf{4}, 14 \href{https://doi.org/10.3390/atoms4020014}%
{(2016)}.

\bibitem{c8} B. Dubetsky, Two approaches in the theory of atom
interferometry, arXiv:\href{https://doi.org/10.48550/arXiv.1701.07909}{%
1701.07909 [physics.atom-ph]}

\bibitem{c9} E. Giese, W. Zeller, S. Kleinert, M. Meister, V. Tamma, A.
Roura, W.P. Schleich, The interface of gravity and quantum mechanics
illuminated by Wigner phase space Atom Interferometry. In Proceedings of the
International School of Physics \textquotedblleft Enrico Fermi"; Tino, G.N.,
Kasevich, M., Eds.; IOS Press: Amsterdam, The Netherlands, \textbf{188}, 171 
\href{https://doi.org/10.3254/978-1-61499-448-0-171}{ (2014)}.

\bibitem{c10} M. Kasevich, S. Chu, Atomic interferometry using stimulated
Raman transitions, Phys. Rev. Lett., \textbf{67}, 181 \href{https://doi.org/10.1103/PhysRevLett.67.181}%
{(1991).}

\bibitem{c11} R. P. Feynman and A. R. Hibbs, Quantum Mechanics and Path
lntegrals (McGraw-Hill, New York, 1965).

\bibitem{c12} P. Storey and C. Cohen-Tannoudji, The Feynman path integral
approach to atomic interferometry. A tutorial, J. Phys. II \textbf{4},1999 
\href{https://doi.org/10.1051/jp2:1994103}{(1994)}.

\bibitem{c12.1} P. Wolf and P. Tourrenc, Gravimetry using atom
interferometers: Some systematic effects, Phys. Lett. A \textbf{251}, 241 
\href{https://doi.org/10.1016/S0375-9601(98)00881-0}{(1999)}.

\bibitem{c13} A. Peters, K. Y. Chung and S. Chu, High-precision gravity
measurements using atom interferometry, Metrologia \textbf{38}, 25 \href{https://doi.org/10.1088/0026-1394/38/1/4}%
{(2001)}.

\bibitem{c14} H. M\"{u}ller, S.-w. Chiow, S. Herrmann, S. Chu, and K.-Y.
Chung, Atom-Interferometry Tests of the Isotropy of Post-Newtonian Gravity,
Phys. Rev. Lett. \textbf{100}, 031101 \href{https://doi.org/10.1103/PhysRevLett.100.031101}%
{(2008)}.

\bibitem{c15} K. Bongs, R. Launay, and M. A. Kasevich, High-order inertial
phase shifts for time-domain atom interferometers, Appl. Phys. \textbf{84},
599 \href{https://doi.org/10.1007/s00340-006-2397-5}{(2006)}.

\bibitem{c16} W. P. Schleich, D. M. Greenberger and E. M Rasel, A
representation-free description of the Kasevich--Chu interferometer: a
resolution of the redshift controversy, New J. Phys. \textbf{15}, 013007 
\href{https://doi.org/10.1088/1367-2630/15/1/013007}{(2013)}.

\bibitem{c16.1} Chris Overstreet, Peter Asenbaum, Mark A. Kasevich,
Physically significant phase shifts in matter-wave interferome, Am J Physics 
\textbf{89}, 324 \href{https://doi.org/10.1119/10.0002638}{(2021)}.

\bibitem{c17} A. Peters, High precision gravity measurements using atom
interferometry, Thesis, Stanford University (1998).

\bibitem{c18} T. L. Gustavson, Precision rotation sensing using atom
interferometry, Thesis, Stanford University (2000).

\bibitem{c19} J. B. Fixler, Atom Interferometer-Based Gravity Gradiometer
Measurements, Thesis, Stanford University (2003)

\bibitem{c20} G. Lamporesi, Determination of the gravitational constant by
atom interferometry, Thesis, University of Firenze (2006).

\bibitem{c21} K. Takase, Precision rotation rate measurments with a mobile
atom interferometer, Thesis, Stanford University (2008).

\bibitem{c22} G. Rosi, Precision gravity measurements with atom
interferometry, Thesis, University of Pisa (2012).

\bibitem{c23} J. M. Hogan, D. M. S. Johnson, and M. A. Kasevich, Lightpulse
atom interferometry, in Proceedings of the International School of Physics
\textquotedblleft Enrico Fermi\textquotedblright\ on Atom Optics and Space
Physics, edited by E. Arimondo, W. Ertmer, W. Schleich, and E. Rasel (IOS
Press, Amsterdam, Netherlands, 2007), \textbf{168}, pp. 411--447.

\bibitem{c23.1} A. Bertoldi, F. Minardi, M. Prevedelli, Phase shift in atom
interferometers: corrections for non-quadratic Lagrangians and
finite-duration laser pulses, Phys. Rev. A. \textbf{99}, 033619 \href{https://doi.org/10.1103/PhysRevA.99.033619}%
{(2019)}.

\bibitem{c24} C. Ufrecht and E. Giese, Perturbative operator approach to
high-precision light-pulse atom interferometry, Phys. Rev. A \textbf{101},
053615 \href{https://doi.org/10.1103/PhysRevA.101.053615}{(2020)}.

\bibitem{c25} B. Dubetsky, Comment on "Perturbative operator approach to
high-precision light-pulse atom interferometry," Phys. Rev. A \textbf{102},
027301 \href{https://doi.org/10.1103/PhysRevA.102.027301}{(2020)}.

\bibitem{c26} J. Glick, T. Kovachy, Feynman Diagrams for Matter Wave
Interferometry, \href{https://doi.org/10.48550/arXiv.2407.11446}{%
arXiv:2407.11446 [physics.atom-ph]}.

\bibitem{c27} S. Kleinerta, E. Kajaria, A. Rouraa, W. P. Schleich,
Representation-free description of light-pulse atom interferometry including
non-inertial effects, Phys. Rept. \textbf{605}, 1 \href{https://doi.org/10.1016/j.physrep.2015.09.004}%
{(2015)}.

\bibitem{c28} G. L. Kotkin and V. G. Serbo, Collection of Problems in
Classical Mechanics. Oxford, New York: Pergamon Press(1971), problem 6.23.

\bibitem{c29} M. A. Kasevich and B. Dubetsky, Kinematic Sensors Employing
Atom Interferometer Phases. US Patent 7, 317, 184. (2005).

\bibitem{c30} M. A. Kasevich and B. Dubetsky, The phase of an \ atom
interferometer as a direct source for precise navigation. Private
communication (2008).

\bibitem{c31} B. Dubetsky, Local positioning system as a classic alternative
to atomic navigation, J. Navig. \textbf{75}, 273 \href{https://doi.org/10.1017/S037346332200008X}%
{(2022)}.

\bibitem{c31.1} Ch. J. Bord\'{e} Theoretical tools for atom optics and
interferometry C. R. Acad. Sci., Paris, Ser IV, Phys. Astrophys. 2, 509 
\href{https://doi.org/10.1016/S1296-2147(01)01186-6}{(2001)}.

\bibitem{c32} L. D. Landau and E. M. Lifshitz, Mechanics, 3rd Edition,
Butterworth-Heinenann, Oxford, 2000, p. 127

\bibitem{c32.1} Propagators refer to differential equation propagators, as
in textbook by R. S. Palais, R. A. Palais, Differential Equations,
Mechanics, and Computation, American . Society, Providence, Rhode Island
Institute for Advanced Study, Princeton, New Jersey, 2009. They differ from
Quantum Mechanics propagators.

\bibitem{c33} M. Kasevich, A. Weis, E. Riis, K. Moler, S. Kasapi, S. Chu,
Atomic velosity selection using stimulated raman transitions, Phys. Rev.
Lett. \textbf{66}, 2297 \href{https://doi.org/10.1103/PhysRevLett.66.2297}{%
(1991)}.

\bibitem{c34} M. A. Kasevich and B. Dubetsky, Kinematic sensors employing
atom interferometer phases, US Patent 7,317,184 (8 January 2008).

\bibitem{c35} M. A. Kasevich and B. Dubetsky, The phase of an atom
interferometer as a direct source for precise navigation, private
communication (2009).

\bibitem{c36} B. Dubetsky, Comment on "Perturbative operator approach to
high-precision light-pulse atom interferometry," Phys. Rev. A \textbf{102},
027301 \href{https://doi.org/10.1103/PhysRevA.102.027301}{(2020)}.

\bibitem{c37} C. Jekeli Inertial Navigation System with Geodetic
Applications, Walter de Gruyter, Berlin, New York (2001)..

\bibitem{c38} G Varoquaux, R A Nyman, R Geiger1, P Cheinet, A Landragin and
P Bouyer, How to estimate the differential acceleration in a two-species
atom interferometer to test the equivalence principle, New J. Phys. \textbf{%
11}, 113010 \href{https://doi.org/10.1088/1367-2630/11/11/113010}{(2009)}.

\bibitem{c39} A. Peters, K. Y. Chung, and S. Chu, Measurement of
gravitational acceleration by dropping atoms, Nature, \textbf{400}, 849 
\href{https://doi.org/10.1038/23655}{(1999)}.

\bibitem{c40} P. Asenbaum, C. Overstreet, T. Kovachy, D. D. Brown, J. M.
Hogan, and M. A. Kasevich, Phase shift in an atom interferometer due to
spacetime curvature across its wave function, Phys. Rev. Lett. \textbf{118},
183602 \href{https://doi.org/10.1103/PhysRevLett.118.183602}{(2017)}.

\bibitem{c41} B. Dubetsky, Sequential large momentum transfer exploiting
rectangular Raman pulses, Phys. Rev. A \textbf{108}, 063308 \href{https://doi.org/10.1103/PhysRevA.108.063308}%
{(2023)}.

\bibitem{c42} P. Wolf, P. Tourrence, Gravimetry using atom interferometers:
some systematic effects. Phys. Lett. A \textbf{251}, 241 \href{https://doi.org/10.1016/S0375-9601(98)00881-0}%
{(1999)}.

\bibitem{c42.1} It should be noted that in the atomic navigator, the dead
time $T_{1}$ may be nonzero, but it must definitely be a multiple of the
interrogation time T so that the 1st and 2nd Raman pulses in a given loop
coincide in time with the 2nd and 3rd pulses in the previous loop.

\bibitem{c43} B. Young, A. Black, M. Boyd, B. Dubetsky, T. Gustavson, L.
Hollberg, M. Kasevich, T. Loftus, M. Matthews, J. Pease, F. Rolle, P.
Studt,T. Tran, A Vitouchkine, and A. Zorn, Cold atom inertial sensors for
precision navigation. Joint Navigation Conference, Colorado Springs,
Colorado, USA, (2011).

\bibitem{c44} M. J. Snadden, J. M. McGuirk, P. Bouyer, K. G. Haritos, and M.
A. Kasevich, Measurement of the Earth's Gravity Gradient with an Atom
Interferometer-Based Gravity Gradiometer, Phys. Rev. Lett. \textbf{81}, 971 
\href{https://doi.org/10.1103/PhysRevLett.81.971}{(1998)}.

\bibitem{c45} J. B. Fixler, G. T. Foster, J. M. McGuirk, M. A. Kasevich,
Atom Interferometer Measurement of the Newtonian Constant of Gravity,
Science \textbf{315}, 74 \href{https://www.science.org/doi/10.1126/science.1135459}%
{(2007)}.

\bibitem{c46} S. Dimopoulos, P. W. Graham, J. M. Hogan, M. A. Kasevich, S.
Rajendran, Gravitational Wave Detection with Atom Interferometry, Phys.
Lett. B \textbf{678}, 37 \href{https://doi.org/10.1016/j.physletb.2009.06.011}%
{(20*09)}.

\bibitem{c47} P. Asenbaum, C. Overstreet, M. Kim, J. Curti, and M.
A.Kasevich, Atom-interferometric test of the equivalence principle at the 10$%
^{\text{-12}}$ level, Phys. Rev. Lett. \textbf{125}, 191101 \href{https://doi.org/10.1103/PhysRevLett.125.191101}%
{(2020)}.

\bibitem{c48} F. Riehle, Th. Kisters, A. Witte, J. Helmcke, Ch. J. Borde,
Optical Ramsey spectroscopy in a rotating frame: Sagnac effect in a
matter-wave interferometry, Phys. Rev. Lett. \textbf{67}, 177 \href{https://doi.org/10.1103/PhysRevLett.67.177}%
{(1991)}.

\bibitem{c49} A. Lenef, T. D. Hammond, E. T. Smith, M. S. Chapman, R. A.
Rubenstein, and D. E. Pritchard, Rotation Sensing with an Atom
Interferometer, Phys. Rev. Lett. \textbf{78}, 760 \href{https://doi.org/10.1103/PhysRevLett.78.760}%
{(1997)}.

\bibitem{c50} T. L. Gustavson, P. Bouyer, and M. A. Kasevich, Precision
Rotation Measurements with an Atom Interferometer Gyroscope, Phys. Rev.
Lett. \textbf{78}, 2046 \href{https://doi.org/10.1103/PhysRevLett.78.2046}{%
(1997)}.

\bibitem{c51} B. Dubetsky, Comment on "Perturbative operator approach to
high-precision light-pulse atom interferometry," Phys. Rev. A \textbf{102},
027301 \href{https://doi.org/10.1103/PhysRevA.102.027301}{(2020)}.

\bibitem{c52} J. F. Clauser, Ultra-high sensitivity accelerometers and
gyroscopes using neutral atom matter-wave interferometry, in Proceedings of
the International Workshop on Matter Wave Interferometry in the Light of
Schrodinger's Wave Mechanics, edited by G. Badurek, H. Rauch, and A.
Zeilinger, 1987; Physica B \& C \textbf{151}, 262 \href{https://doi.org/10.1016/0378-4363(88)90176-3}%
{(1988)}.

\bibitem{c53} L. A. Sidorenkov, R. Gautier, M. Altorio, R. Geiger, A.
Landragin, Tailoring multi-loop atom interferometers with adjustable
momentum transfer, Phys. Rev. Lett. \textbf{125}, 213201 \href{https://doi.org/10.1103/PhysRevLett.125.213201}%
{(2020)}.

\bibitem{c54} J. K. Stockton, K. Takase, M. A. Kasevich, Absolute geodetic
rotation measurement using atom interferometry. Phys. Rev. Lett. \textbf{107}%
, 133001 \href{https://doi.org/10.1103/PhysRevLett.107.133001}{(2011)}.

\bibitem{c55} I. Dutta, D. Savoie, B. Fang, B. Venon, C. L. Garrido Alzar,
R. Geiger, and A. Landragin, Continuous Cold-Atom Inertial Sensor with 1
nrad/sec Rotation Stability, Phys. Rev. Lett. \textbf{116}, 183003 \href{https://doi.org/10.1103/PhysRevLett.116.183003}%
{(2016)}.

\bibitem{c57} B. Canuel, F. Leduc, D. Holleville, A. Gauguet, J. Fils, A.
Virdis, A. Clairon, N. Dimarcq, Ch. J. Borde, A. Landragin, 6-axis inertial
sensor using cold-atom interferometry, Phys. Rev. Lett. \textbf{97}, 010402 
\href{https://doi.org/10.1103/PhysRevLett.97.010402}{(2006)}.

\bibitem{c58} L. R. Jaroszewicz, A. Kurzych, Z. Krajewski, M. Dudek, J. K.
Kowalski 2, K. P. Teisseyre, The Fiber-Optic Rotational Seismograph -
Laboratory Tests and Field Application, Sensors , \textbf{19}, 2699 \href{https://doi.org/10.3390/s19122699}%
{(2019)}.
\end{thebibliography}
\end{document}